\documentclass[journal=jpclcd,manuscript=letter,layout=twocolumn]{achemso}

\usepackage[version=3]{mhchem} 
\usepackage{xcolor}
\usepackage{mciteplus}
\usepackage{chemfig}
\usepackage[font=scriptsize]{caption}
\usepackage{amsmath}
\usepackage{graphicx}

\usepackage{enumitem}

 \usepackage[english]{babel}
 \usepackage{stfloats}

\author{Sung Kwon}
\affiliation[Michigan State University]
{Department of Chemistry, Michigan State University, 48824 East Lansing, MI, United States}
\author{Naga Krishnakanth Katturi}
\affiliation[Michigan State University]
{Department of Chemistry, Michigan State University, 48824 East Lansing, MI, United States}
\author{Bruno I. Moreno}
\affiliation[Universidad de Chile]
{Department of Physics, University of Chile, CEDENNA,  Las Palmeras, 3425, Ñuñoa, Chile}
\author{Carlos Cárdenas}
\affiliation[Universidad de Chile]
{Department of Physics, University of Chile, CEDENNA, Las Palmeras, 3425, Ñuñoa, Chile}

\email{dantus@chemistry.msu.edu}
\email{cardena@uchile.cl}
\author{Marcos Dantus}
\affiliation[Michigan State University]
{Department of Chemistry, Michigan State University, 48824 East Lansing, MI, United States}
\alsoaffiliation[Michigan State University]
{Department of Physics and Astronomy, Michigan State University, 48824 East Lansing, MI, United States}
\alsoaffiliation[Michigan State University] 
{Department of Electric and Computer Engineering, Michigan State University, 48824 East Lansing, MI, United States}
\email{dantus@chemistry.msu.edu}

\title{Ultrafast cation-dication dynamics in ammonia borane: \ce{H}-migration to roaming \ce{H2}
and reduced \ce{H3+} formation under strong-field ionization}

\abbreviations{MS}


\begin{document}
\maketitle
\section{TOC Graphic}
\begin{figure}[t]
 \centering

 \includegraphics[width=3.25in,height=1.75in]{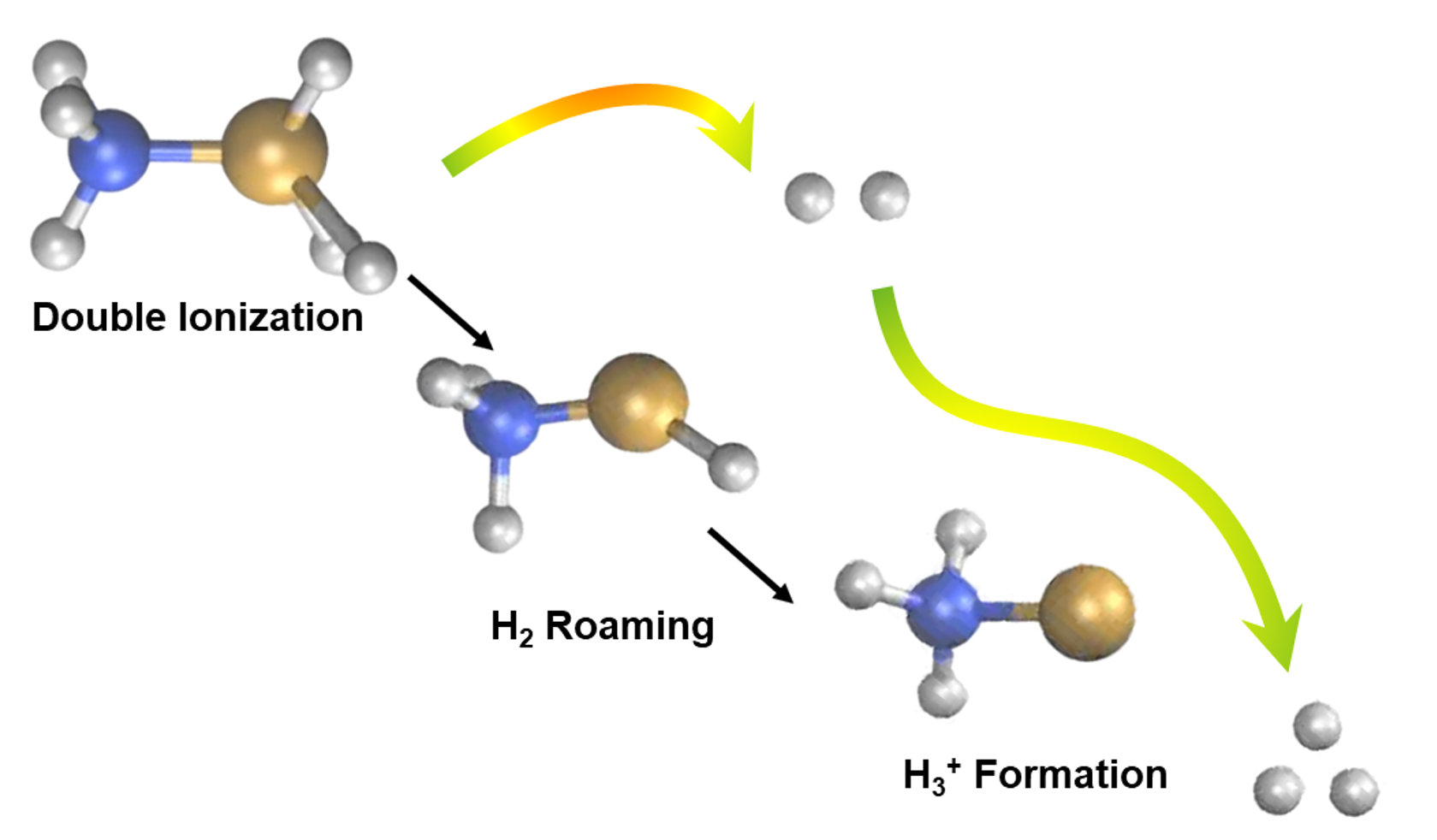}
  \label{TOC}
 \end{figure}

\begin{abstract}
We report a femtosecond time-resolved strong-field study of ammonia borane (AB, \ce{BH3NH3}) following both single and double ionization, revealing ultrafast fragmentation dynamics and hydrogen release. Mass spectrometry, combined with fragment correlation analysis and \textit{ab initio} molecular dynamics simulations, is used to identify the molecular origin of the neutral and ionic products. Singly ionized AB produces neutral \ce{H} and \ce{H2}, while doubly ionized AB produces neutral H and \ce{H2} along with  \ce{H+}, \ce{H2+}, and \ce{H3+}, all within 1 ps. Electronic-structure calculations show that  \ce{H}, \ce{H+}, \ce{H2}, \ce{H2+}, and \ce{H3+} originate predominantly from hydrogen atoms bound to the boron center and that their formation proceeds through hydrogen migration and, in some channels, neutral \ce{H2} roaming. The calculations further indicate that the dication meets the structural and energetic requirements for neutral \ce{H2} release, a prerequisite for forming astrochemically relevant \ce{H3+}. However, the large adiabatic relaxation energy causes most roaming \ce{H2} to dissociate before proton abstraction, suppressing \ce{H3+} formation. These results provide new insight into dissociative ionization pathways in hydrogen-rich molecules, extend mechanistic principles developed for halogenated alkanes to ammonia borane, and suggest implications for hydrogen-release chemistry in ammonia-borane-based storage materials.

\end{abstract}

\section{Introduction}
Ammonia borane (AB, \ce{BH3NH3}) has attracted significant attention as a chemical hydrogen storage material due to its exceptionally high hydrogen content, 19.6 wt$\%$, which exceeds that of many conventional storage materials.\cite{chen2008recent,graetz2009new,sutton2011regeneration,DOE_H2Storage_ExecSummaries}
The dehydrogenation process of AB is a key area of research, given its relevance to hydrogen storage applications. Experimental studies have shown that thermal decomposition releases hydrogen in a stepwise manner, forming intermediates such as polyaminoborane and borazine.\cite{miranda2007ab}
While most studies have focused on condensed-phase or thermal dehydrogenation behavior of AB, little is known about its gas-phase dynamics under ionizing conditions. The electron-ionization mass spectrum of AB is not available in existing databases, and little is known about its fragmentation behavior upon ionization. In this study, we examine the ultrafast, time-resolved dissociative dynamics of AB following strong-field double ionization. \par

The chemical bonding in AB involves a dative (coordinate covalent) bond, in which the nitrogen atom donates a lone pair of electrons to the electron-deficient boron atom. The structural properties of AB have been extensively studied to understand its hydrogen storage capabilities. Photoemission and X-ray absorption studies have been conducted to probe its valence and core electronic states, shedding light on the nature of bonding and electronic transitions within the molecule.\cite{sa2023photoemission} Theoretical investigations using ab initio molecular dynamics and metadynamics have provided insights into the initial stages of dehydrogenation, suggesting the formation of intermediates like ammonia diborane, which can lead to autocatalytic hydrogen production cycles.\cite{rizzi2019onset} 
Recent studies have also explored the potential of ammonia borane in carbon dioxide capture and conversion.\cite{kumar2021reduction,castilla2022ammonia} 
\par
 Several molecules have been shown to release \ce{H3+} upon double ionization. For example, Eland demonstrated that \ce{H3+} can be produced from a range of organic molecules through collisions with high-energy electrons or excitation by 30.4 nm photons.\cite{Eland1996TheOrigin} Similarly, Yamanouchi and co-workers (2005-2006) reported \ce{H3+} ejection from gas-phase methanol under intense 800 nm laser fields.\cite{furukawa2005ejection,okino2006ejection} Complementing these studies, our group has been investigating ultrafast, far-from-equilibrium chemical processes initiated by interactions between molecules and secondary electrons.\cite{dantus2024tracking,dantus2024ultrafast} In particular, we have focused on the formation dynamics and mechanisms of \ce{H3+} generation from various organic molecules following strong-field ionization.\cite{ekanayake2017mechanisms, ekanayake2018h2, ekanayake2018substituent, michie2019quantum, kwon2023mechanism, stamm2025factors} Using EUV excitation, Strasser and collaborators have also investigated \ce{H3+} formation dynamics in several of these same systems, including methanol and ethanol.\cite{gope2022inverse,gope2023sequential} Despite the prevalence of organic precursors in these studies, only a few non-organic molecules, such as AB examined here, and \ce{GeH4}, have been shown to generate \ce{H3+} upon ionization.\cite{Eland1996TheOrigin} The production of \ce{H3+} is significant beyond laboratory conditions, it is a key ion in interstellar chemistry, acting as a catalytic proton donor,\cite{herbst1973formation,oka2013interstellar} and has recently been proposed as an indirect probe for dark matter.\cite{luque2025anomalous} Interestingly, AB is isoelectronic with ethane, a molecule known to readily form \ce{H3+} via strong-field dissociation.\cite{burrows1979studies,hoshina2011metastable,kraus2011unusual,schirmel2013formation,li2020control} This structural similarity raises questions about the potential of AB to follow analogous dissociation pathways in the gas phase.
\par

\section{Methods}
The ultrafast disruptive-probing method used to monitor all reaction pathways following strong-field ionization has been detailed previously.\cite{jochim2022ultrafast} In this case, a Ti:sapphire laser centered at 795 nm, delivering 65 fs pulses at 1 kHz as measured by in-situ autocorrelation, served as the ionization source. Ammonia borane (95\% purity, Sigma-Aldrich) was used without further purification. Laser pulses were focused into a Wiley–McLaren time-of-flight (TOF) mass spectrometer using a 200 mm lens, with polarization aligned parallel to the TOF axis. Each pulse was split into a strong pump pulse to ionize the molecule and a weak probe pulse to disrupt product formation. Laser intensities were calibrated by measuring the \ce{Ar^{2+}}/\ce{Ar+} ion ratio.\cite{guo1998single}  
Ammonia borane vapor was introduced into the TOF chamber as an effusive beam after a freeze-pump-thaw-cycle. Room-temperature sublimation provided sufficient vapor to obtain the experimental data. During data acquisition, the chamber pressure was held at 1 × 10$^{-5}$ Torr. The baseline vacuum pressure was 9 × 10$^{-8}$ Torr and returned to this level within a few seconds of closing the valve. Ion signals were digitized with a 1 GHz oscilloscope (LeCroy WaveRunner 610Zi). The TOF was set to detect cations, so only positively charged ions appear in the mass spectra. To account for the natural isotopic distribution of boron (80\% $^{11}$B and 20\% $^{10}$B), peaks containing boron, except the molecular ion, were corrected by 20\% to minimize isotope-related contributions. Additionally, contributions from air in the spectrum were removed, using the ratio of ionized \ce{N2+} to \ce{O2+} in air under identical laser conditions to subtract the \ce{N2+} contribution from the m/q 28 signal. 

The correlation coefficient analysis with shot-to-shot correction was carried out to account for laser intensity fluctuations. A total of N = 39,660 single-shot mass spectra of AB were acquired and analyzed for this purpose. From the same dataset, the kinetic energy release (KER) distributions for \ce{H+}, \ce{H2+}, and \ce{H3+} were also extracted. 

Electronic structure calculations, optimization of the ground state and dication geometries of AB, adiabatic relaxation energies, and \ce{H2} dissociation energies were carried out at the CCSD(T)\cite{raghavachari1989fifth, bartlett1990non,piecuch2002efficient,piecuch2005renormalized}/aug-cc-pVDZ\cite{dunning1989gaussian,kendall1992electron} level of theory using GAMESS 2019.R1.\cite{GAMESS} The ionization potential calculations for the monocation were done at the CCSD\cite{purvis1982full,cullen1982linked,piecuch2002efficient,piecuch2005renormalized}/cc-pVTZ\cite{dunning1989gaussian,woon1993gaussian} and dication were done at the DIP-EOMCC(4h-2p)\cite{vcivzek1966correlation,vcivzek1969use,paldus1972correlation,shen2013doubly,shen2014doubly}/cc-pVTZ\cite{dunning1989gaussian,woon1993gaussian} level of theory using the same version of GAMESS. All molecular dynamics simulations were performed using Gaussian 09\cite{gaussian09} evaluating the force with DFT using the $\omega$-B97XD\cite{w97xd} exchange-correlation functional with a 6-311+G(d,p) basis set. To better capture the potential energy of bond breaking processes, a spin broken-symmetry solution was allowed for trajectories. Equations of motion were integrated with Verlet-Velocity and a Hessian-based integrator\cite{hase1999} with step size of 0.5 fs and 0.25 $\sqrt{amu}*bohr$, respectively. For the cation's AIMD trajectories, atomic positions were taken from a molecular dynamics (MD) simulation of the neutral AB in the NVT (conserving the number of particles, volume, and kinetic energy) ensemble at 300~K (with 0~eV of additional energy). The velocities were sampled from a Maxwell–Boltzmann distribution (thermal sampling), with the constraint that the total kinetic energy of the nuclei was scaled to 1, 1.5, or 7.8~eV. For the case labeled as 0~eV, the velocities were obtained directly from the NVE MD simulation (Maxwell–Boltzmann at 300~K). A total of 100 trajectories were generated for each energy level. 

For the dication's AIMD trajectories, two distinct cases were considered. In the first case, a set of 200 trajectories was generated, where the positions and velocities were derived from NVT molecular dynamics simulations of the neutral state (Maxwell–Boltzmann distribution at 300~K). In this scenario, the adiabatic energy is sufficient to promote hydrogen (\ce{H}) release but not enough to induce Coulomb fragmentation of the boron–nitrogen (B–N) bond. To observe B–N bond fragmentation (second case), 4~eV of energy was selectively injected into the normal mode of \ce{BH3NH3^2+} with the largest B–N stretching character, using quasi-classical sampling. This approach was chosen because B–N bond cleavage is a rare event compared to \ce{H} release, requiring thousands of thermally sampled trajectories to observe it. In some trajectories, fragmentation did not lead to bond breaking but rather to additional hydrogen release, producing \ce{BHNH3^2+}. For these cases, we further injected 4~eV into the B–N stretching mode of \ce{BHNH3^2+}, and so on. Only 25 trajectories were run for these higher-energy cases, as increasing the number of trajectories did not significantly affect the observed trends.

\section{Results and Discussion}
\begin{figure}[!h]
 \centering
  \includegraphics [width = \columnwidth]{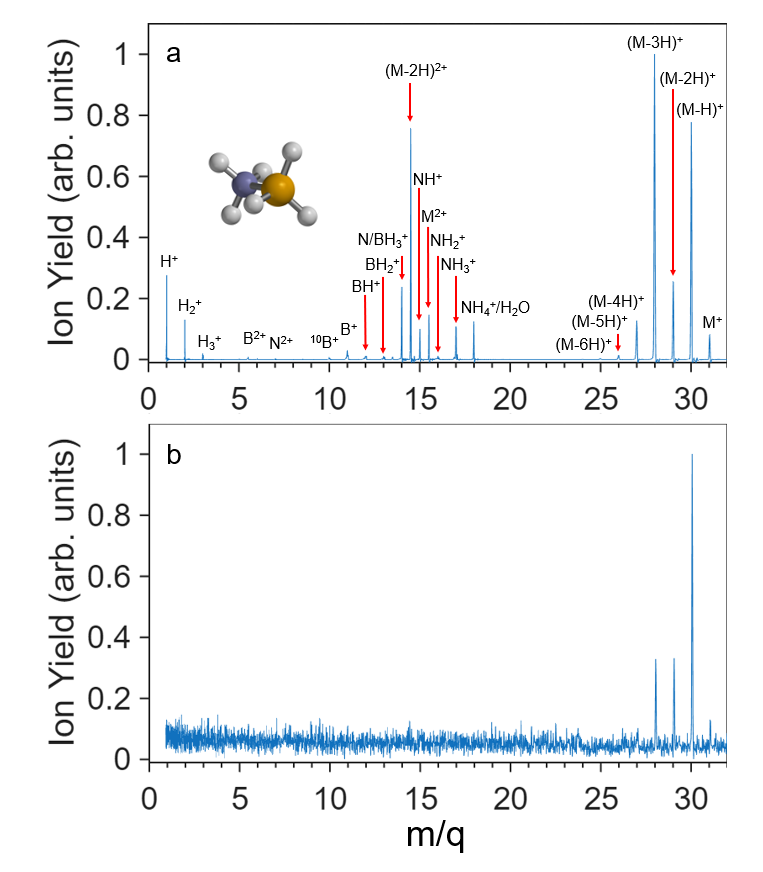}
  \caption{(a) Strong-field ionization mass spectrum of ammonia borane after the difference between two different laser peak intensities of $3.1 \times 10^{14}$ W/cm$^2$ and $2.7 \times 10^{14}$ W/cm$^2$. Fragment peaks are labeled with their sum formulas. (b) Low-intensity strong-field ionization mass spectrum of ammonia borane taken at an intensity of $1.3 \times 10^{14}$ W/cm$^2$. Both spectra are normalized to a maximum intensity of 1. M corresponds to the molecular ion.}
  \label{Fig1}
 \end{figure}

The experimental high-intensity selected strong-field ionization mass spectrum of \ce{BH3NH3} resulting from the difference between two different laser peak intensities of $3.1 \times 10^{14}$ W/cm$^2$ and $2.7 \times 10^{14}$ W/cm$^2$ is shown in Figure \ref{Fig1}a. This difference intensity minimizes  contributions from the lower-intensity regions of the Gaussian focal-volume intensity distribution.\cite{posthumus2004dynamics,wang2005disentangling,wiese2019strong} 
The mass spectrum shows that under strong-field ionization AB can lose all its hydrogen atoms. In addition, dication species are observed at m/q 15.5 (M$^{2+}$), 14.5 (M–2H$^{2+}$), 13.5 (M–4H$^{2+}$), and 5.5 (B$^{2+}$). All smaller fragment ions, beginning with \ce{NH3+}, exhibit Coulomb-explosion signatures in the time-of-flight spectrum, characterized by distinct forward and backward peaks, except for those at m/q 15.5, 15, 14.5, and 13.5, which appear as single peaks and are assigned as the molecular dication or the molecular dication with the loss of one to three hydrogen atoms. We note that the m/q 15 channel may also contain a contribution from \ce{NH+}. The low-intensity $1.3 \times 10^{14}$ W/cm$^2$ ionization mass spectrum is shown in Figure \ref{Fig1}b. This spectrum shows that even at low intensities, AB loses one or more H atoms. The high yield of m/q 30, corresponding to loss of a single hydrogen atom, is attributed to the small energy gap between the adiabatic ionization potential of AB (9.6 eV) and the threshold for the first H-atom loss (10.0 eV).\cite{schleier2022ammonia}  Hydrogen loss is quantified in Tables 1 and 2 from the integrated peak areas for the monocation and dication, respectively, and both tables show that hydrogen loss is frequent following both single and double ionization.

\begin{table}[h]
  \centering
  \caption{The experimental integrated areas of m/q fragments corresponding to H loss normalized by the sum of the total integrated area of all peaks at an intensity of $1.3 \times 10^{14}$ W/cm$^2$.}
  \begin{tabular}{c c}
    \hline
    m/q & Area (\% of Monocations) \\
    \hline
    31 & 7.5 \\
    30 & 38.6 \\
    29 & 13.9 \\
    28 & 32.8 \\
    27 & 7.3 \\
    \hline
    \label{tab1}
  \end{tabular}
\end{table}

\begin{table}[h]
  \centering
  \caption{The experimental areas of m/q fragments corresponding to loss of 2H(s) normalized to the total area of dication peaks at an intensity of $3.1 \times 10^{14}$ W/cm$^2$.}
  \begin{tabular}{c c}
    \hline
    m/q & Area (\% of Dications) \\
    \hline
    15.5 & 15.3 \\
    14.5 & 80.2 \\
    13.5 & 1.4 \\
    \hline
    \label{tab2}
  \end{tabular}
\end{table}

\begin{figure}[h!]
 \centering
 \includegraphics[width=\columnwidth]{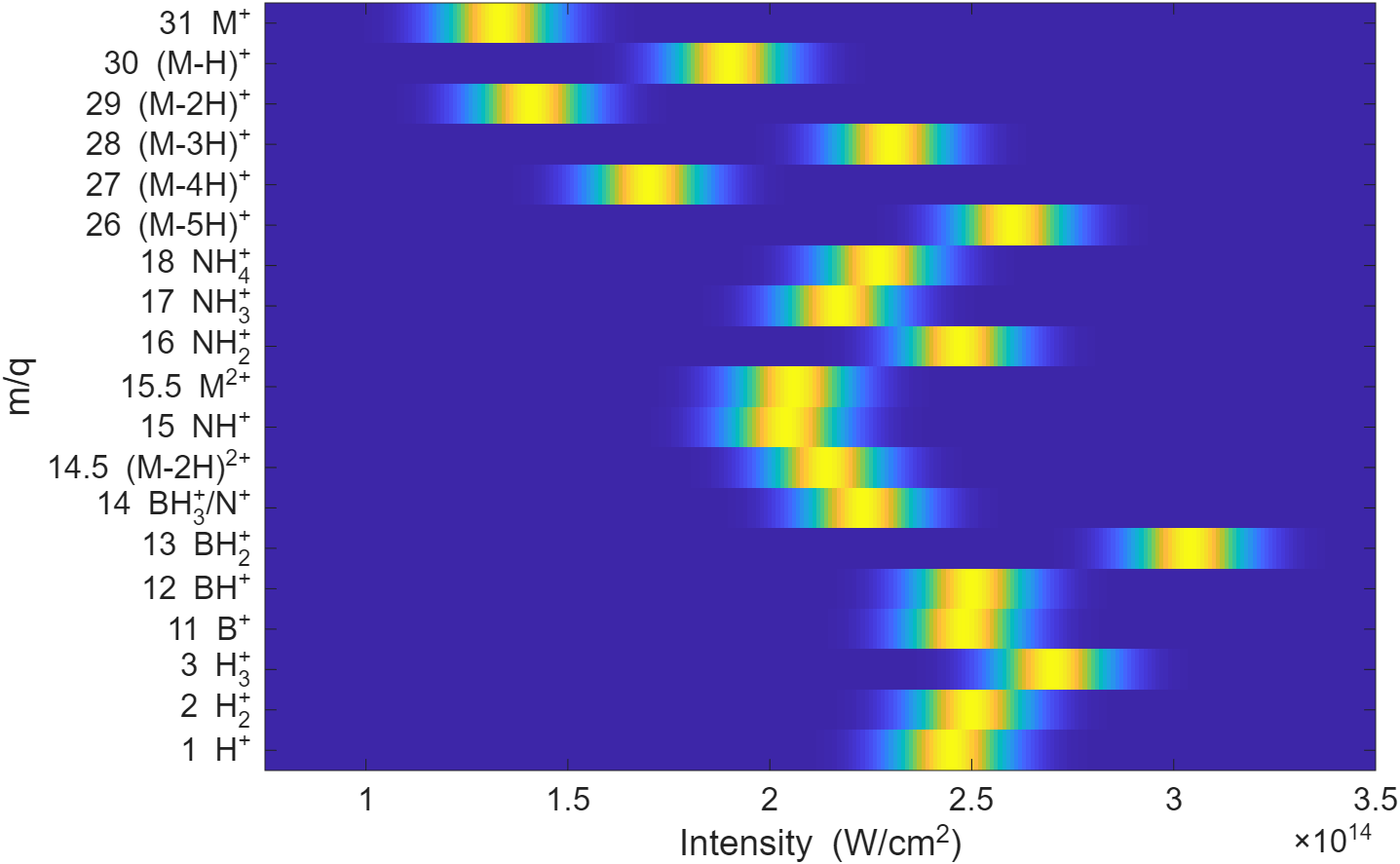}
 \caption{Heat map of the major AB fragment yields as a function of calibrated laser intensity. The appearance-intensity thresholds, which are proportional to the corresponding appearance energies, were extracted using the procedure described in the Supporting Information (Figure S1). The width of each onset feature reflects the uncertainty in the threshold determination and thus provides a measure of confidence.}
 \label{Fig2}
\end{figure}
Dissociative ionization was further examined through calibrated\cite{guo1998single} laser power dependence in Figure \ref{Fig2}, where each successive data point corresponds to a higher intensity. For each fragment, we compute the numerical difference in integrated yield between adjacent points (i.e. point n+1 minus point n), which suppresses the focal-volume contribution. The resulting difference trace is then fit with an error function. The appearance of each fragment was determined by the laser peak intensity at which its yield exceeded 10\% of the maximum (see Figure S1). 

When photoionization occurs within a single optical cycle or faster, the liberated electron gains kinetic energy proportional to the peak intensity. The ponderomotive energy, which scales linearly with intensity, governs the electron’s motion, driving it away from the ion and back toward it as the laser field reverses.\cite{corkum1993plasma} In general, the intensity required for ion formation can be directly related to the fragment’s appearance energy (AE). Upon recollision, the electron transfers energy to the molecule, inducing ionization, fragmentation, and in some cases, double ionization. The calculated single and double vertical ionization potentials for AB were calculated to be 11.9 eV and 31.1 eV, vide infra at the CCSD\cite{purvis1982full,cullen1982linked}/cc-pVTZ\cite{dunning1989gaussian,woon1993gaussian} and DIP-EOMCC(4h-2p)\cite{vcivzek1966correlation,vcivzek1969use,paldus1972correlation,shen2013doubly,shen2014doubly}/cc-pVTZ\cite{dunning1989gaussian,woon1993gaussian} respectively. In general, we see that losing pairs of hydrogen atoms requires less energy compared to single or triple hydrogen loss. Unlike most hydrocarbons,\cite{lozovoy2008control} \ce{H+} loss from AB is observed only when the laser intensity is high enough to cause double ionization. Note that fragments resulting from breaking the B-N bond (m/q 18, 17, 16, 14, 13, 12, and 11) exceed the appearance energy of the dication.
\par

\begin{figure*}[!h] 
 \centering
 \includegraphics[width=\textwidth]{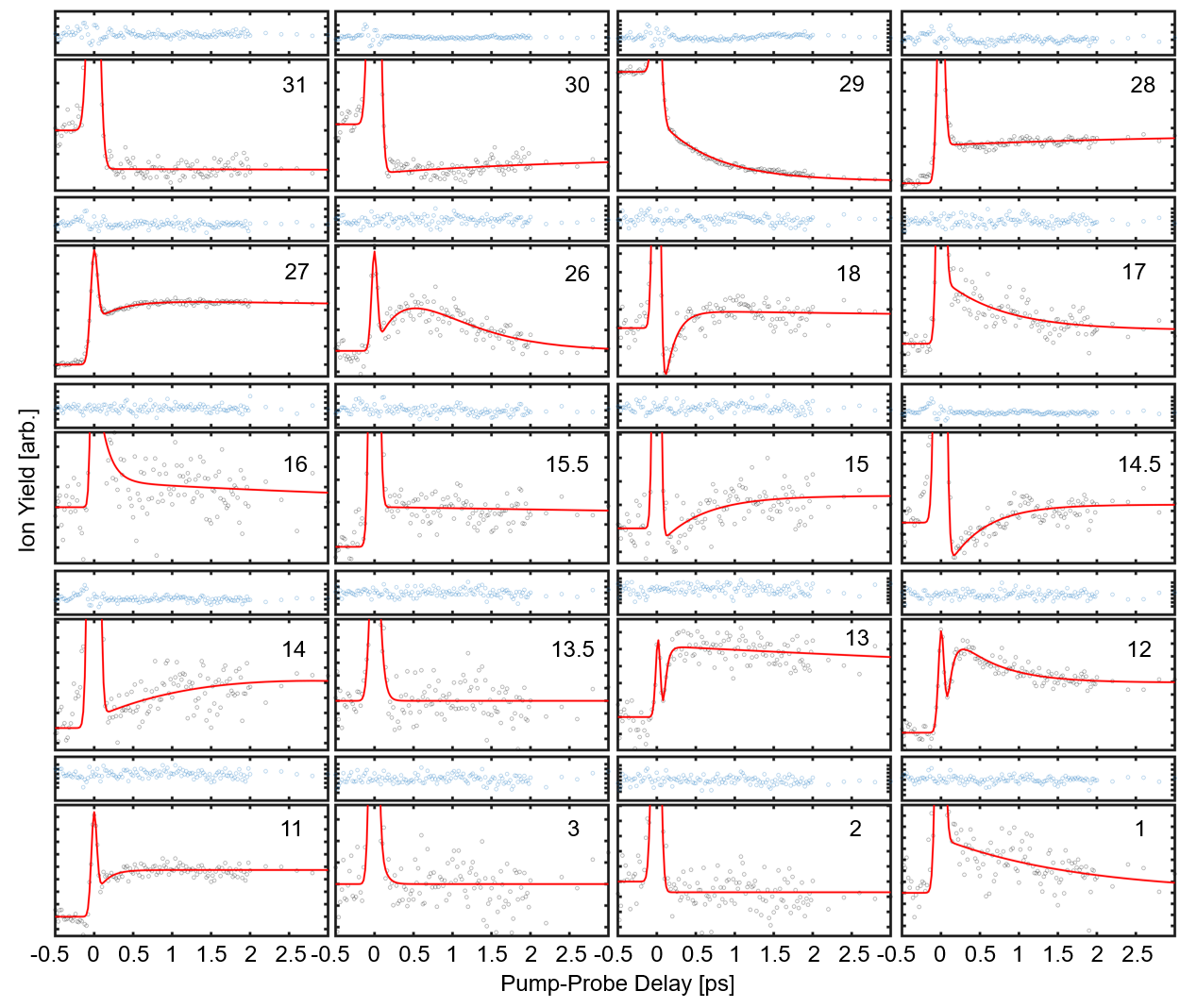}
 \caption{Ion yield as a function of pump-probe delay for all major m/q with the residual plotted above. The time-resolved data was acquired at a pump intensity of $3.1 \times 10^{14}$~W/cm$^2$ and probe intensity of $4.4 \times 10^{13}$~W/cm$^2$.}
 \label{Fig3}
\end{figure*}

The ultrafast formation dynamics of all ionic species were simultaneously measured using disruptive probing, a technique where an intense pump pulse ionizes the molecule and a time-delayed weak probe pulse disrupts its fragmentation.\cite{jochim2022ultrafast} The probe pulse by itself causes no ionization. The resulting ion yields as a function of pump-probe delay time enable the simultaneous tracking of all fragmentation pathways and products.\cite{jochim2022ultrafast} Figure \ref{Fig3} shows the ion yield as a function of pump-probe delay for all major fragments of AB. 
\begin{table*}[h!]
  \centering
  \begin{tabular}{c c c c c c c}
    \hline
    m/q & $\tau_1$ (fs) & $\tau_2$ (fs) & $\tau_3$ (fs) & $a_1$ & $a_2$ & $a_3$ \\
    \hline
    31  & $92 \pm 10$ & offset & - & 0.01 & -0.004 & - \\
    30  & $250 \pm 180$ & $10000 \pm 530$ & - & -0.002 & 0.0008 & - \\
    29  & $340 \pm 65$ & offset & - & 0.01 & -0.01 & - \\
    28  & $3500 \pm 1300$ & offset & - & -0.004 & 0.01 & - \\
    27  & $340 \pm 100$ & offset & - & -0.001 & 0.004 & - \\
    26  & $175 \pm 10$ & $810 \pm 35$ & - & 0.02 & -0.03 & - \\
    18  & $140 \pm 47$ & offset & - & -0.02 & 0.002 & - \\
    17  & $720 \pm 75$ & offset & - & 0.005 & 0.003 & - \\
    16  & $130 \pm 20$ & $5100 \pm 2300 $ & - & 0.01 & .003 & - \\
    15.5 & offset & - & - & 0.004 & - & - \\
    15  & $640\pm 78$ & offset & - & -0.005 & 0.003 & - \\
    14.5 & $480\pm 55$ & offset & - & -0.006 & 0.002 & - \\
    14  & offset & - & - & 0.002 & - & - \\
    13.5 & $40 \pm 32$ & - & - & 0.4 & - & - \\
    13  & $40 \pm 40$ & offset & - & -0.1 & 0.02 & - \\
    12  & $80 \pm 44$ & $480 \pm 270 $ & offset & -0.07 & 0.02 & -0.02 \\
    11  & $120 \pm 16$ & offset & - & 0.02 & -0.01 & - \\
     3  & offset & - & - & 0.004 & - & - \\
     2  & $230 \pm 170$ & - & - & 0.002 & - & - \\
     1  & $400 \pm 200$ & - & - & 0.005 & - & - \\
    \hline
  \end{tabular}

  \caption{Exponential fit parameters for all $m/q$ ion dynamics from AB.}
  \label{tab3}
\end{table*}

The time-dependent yield of the selected fragments exhibit a narrow feature at zero time delay, corresponding to the spatial and temporal overlap of the pump and probe pulses, as shown in Figure \ref{Fig3}. At asymptotic pump–probe delays, the ion yields for m/q 31, 30, and 29 remain reduced relative to their negative-delay baselines. In contrast, the yields of all other major fragments either (i) recover to the negative-delay value—indicating that, at long delays, the probe no longer measurably perturbs their formation or depletion—or (ii) increase above the baseline, consistent with probe-induced formation from a long-lived precursor. At early (sub-picosecond) delays, transient depletions or enhancements reflect probe perturbation of the evolving reaction dynamics; fitting these transients provides an effective formation timescale for the corresponding product. Owing to the low signal-to-noise ratio, we could not reliably extract a formation timescale for \ce{H3+}. Loss of two hydrogens leading to m/q 29 displays a depletion with a time constant of $340 \pm 65$~fs and a long offset term, while loss of four hydrogens leading to m/q 27 shows a fast rise with a 340 $\pm$ 100 fs time constant and a long offset term. The similarity between these timescales suggests that both two- and four-hydrogen loss channels proceed through closely related dynamics. On the other hand, the loss of five hydrogen atoms leading to m/q 26 shows a rapid enhancement with a $175 \pm 10$~fs time constant, followed by a slower decay with a $810 \pm 35$~fs time constant. A similar enhancement is observed for m/q 13 and 12, though it occurs more rapidly, with time constants of $40 \pm 40$~fs and $80 \pm 44$~fs, respectively. The formation of \ce{NH4+} (m/q 18) is characterized by a time constant of $140 \pm 47$~fs. This rapid hydrogen migrations from boron to nitrogen occurs on a timescale consistent with previously reported H-transfer dynamics.\cite{de2000femtosecond,stamm2023surprising,minvielle2025isomer}
All the fitting constants are provided in table \ref{tab3}. In certain cases (m/q = 29, 27, and 26), the dynamics were corrected for contamination from $^{10}$B. This correction, applied from high to low m/q, was based on isotope contributions estimated from the $^{11}$B-corrected fragment signal. Specifically, the ion yield of the affected m/q was scaled at each time delay using its corresponding correction factor. For m/q 31, 30, 28, and 26, isotope contributions were less than 3$\%$ and thus not corrected. The dynamics observed under high-intensity conditions closely match those at low intensity (see Figure S2), indicating that fragments in the m/q 26–31 range predominantly arise from monocationic AB, even at high intensity. The dication corresponding to (M-2H)$^{2+}$ exhibits a depletion with a time constant of $230 \pm 43$ fs, consistent with \ce{H2} loss. Dicationic fragments at m/q 15.5 and 13.5 show no discernible time-resolved dynamics. 
\par
\begin{figure}[!h]
 \centering
 \includegraphics[width=\columnwidth]{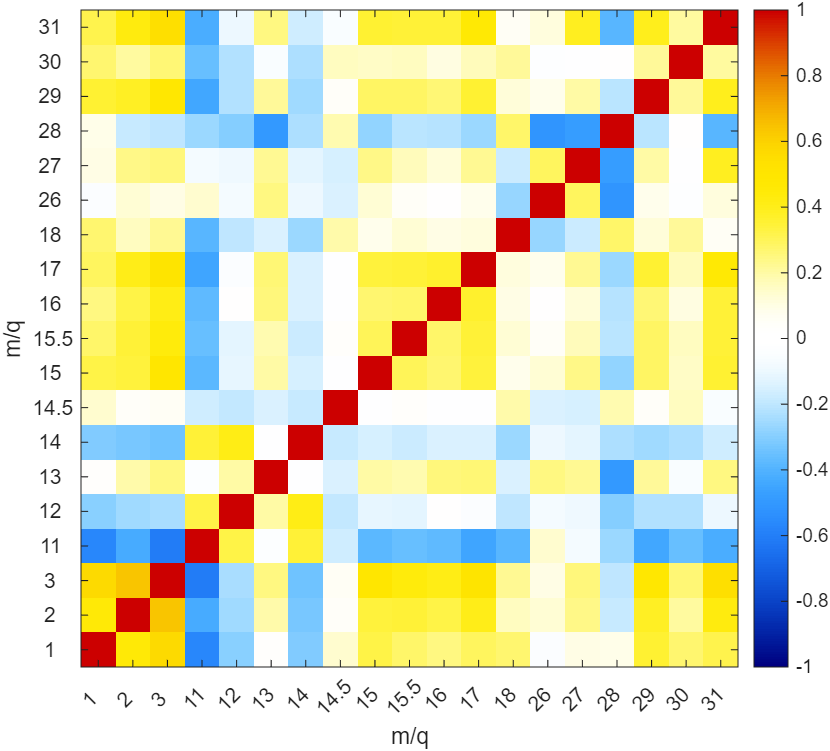}
 \caption{The calculated correlation coefficients from N = 39,660 single-shot mass spectra of AB after shot-to-shot corrections.}
 \label{Fig4}
\end{figure}

Fragment correlation mass spectrometry is a technique to infer common parentage, concerted breakup, or reaction pathways among ions.\cite{frasinski1989covariance,li2024fragment,li2024fragment1} We measured the dissociative ionization of AB fragment ions at an intensity of $3.1 \times 10^{14}$ W/cm$^2$, while keeping the background pressure at $10^{-7}$ Torr. We acquired 39,660 single shot mass spectra and analyzed the correlation coefficients of all the fragments. For m/q 2, 11, 12, 13, 14, 16, and 17, the signals exhibit clear forward–backward splitting (see Figure S3). To isolate contributions from dicationic channels, only the wings of these peaks were used in the correlation analysis. To account for shot-to-shot fluctuations, the ion signals from all fragments were summed and used for normalization. The resulting correlation coefficient matrix is shown in Figure \ref{Fig4}. In the absence of shot-to-shot fluctuation correction, nearly all fragments exhibit spurious positive correlations, as illustrated by the uncorrected matrix in the Supporting Information (see Figure S4).

Analysis of single-shot mass spectra of \ce{AB} following dissociative double ionization provides insight into the origin of \ce{H2} loss and subsequent fragment formation. We observe positive correlation signals between \ce{H+}, \ce{H2+}, and \ce{H3+} and \ce{NH_n^+} (n = 1–3). These correlations indicate that the hydrogens forming \ce{H+}, \ce{H2+}, and \ce{H3+} through \ce{H2} loss predominantly originate from the boron end of the molecule.
\par
The time-of-flight arrival of \ce{H+}, \ce{H2+}, and \ce{H3+}, can be further analyzed to determine their kinetic energy release (KER). In our experiments, the laser polarization was oriented parallel to the time-of-flight (TOF) axis. Because ammonia borane has a strong permanent dipole moment, molecules with their dipole aligned along the laser polarization are expected to ionize most efficiently. As a result, we anticipate strongly polarized forward--backward ejection of the \ce{H_n^+} fragments along the polarization (TOF) axis.

The fragment-ion kinetic energy, $E_{\mathrm{ion}}$, was obtained from the time separation between the forward and backward peaks, $\Delta t$, according to
\begin{equation}
E_{\mathrm{ion}} = \frac{q^2 F^2 (\Delta t)^2}{8m},
\end{equation}
where $q$ is the fragment charge, $F$ is the static extraction field, $\Delta t$ is the forward--backward flight-time difference, and $m$ is the fragment mass. To convert this value to the total kinetic energy release (KER) of the two-body breakup, we account for recoil of the complementary fragment with mass $M - m$, where $M$ is the parent-ion mass, using
\begin{equation}
E_{\mathrm{total}} = E_{\mathrm{ion}} \left( 1 + \frac{m}{M - m} \right).
\end{equation}

\begin{figure}[h!]
 \centering
 \includegraphics[width=0.4\textwidth]{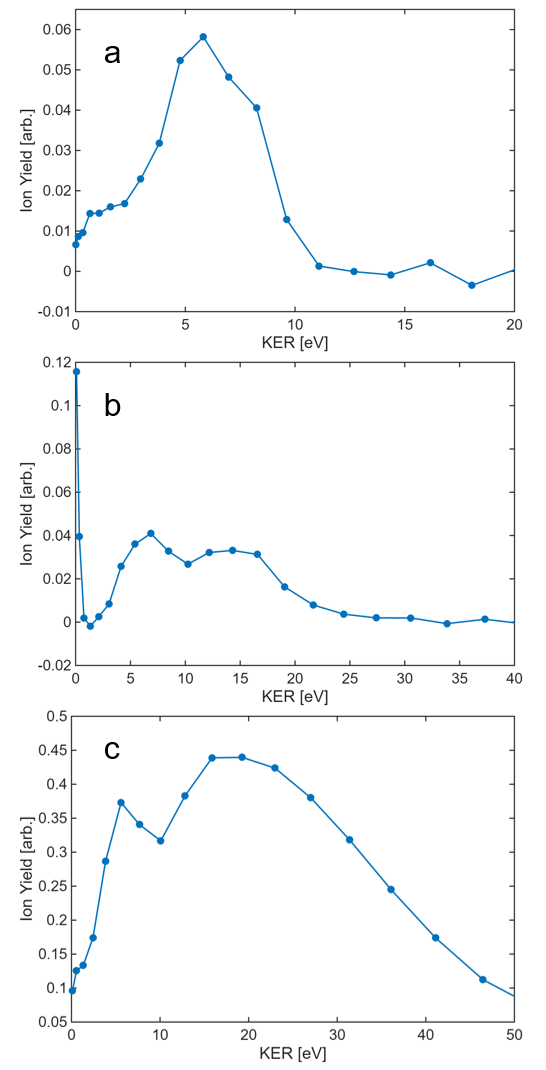}
 \caption{The experimental KER spectra of (a) \ce{H3+}, (b) \ce{H2+}, and (c) \ce{H+}. Blue dots represent the measured data, while the blue line serves as a guide to the eye.}
 \label{Fig5}
\end{figure}

These expressions were used consistently for all reported KER values. As shown in Figure \ref{Fig5}, the KER distributions for \ce{H+}, \ce{H2+}, and \ce{H3+} exhibit prominent peaks at 5.7, 6.1, and 5.8 eV, respectively. These values are consistent with Coulomb explosion following double ionization and are comparable to the KER expected for doubly ionized methanol (5.0 eV if all three \ce{H} atoms originate from carbon and 5.48 eV if one \ce{H} atom originates from oxygen).\cite{ekanayake2017mechanisms} Because KER is highly sensitive to the interatomic distance at the moment of charge separation, even small changes in geometry can shift the peak substantially; for example, the KER for \ce{N2} is 6.5 eV.\cite{strasser2023single} In addition to the low-KER peaks, \ce{H2+} and \ce{H+} show higher-energy peaks at 13 and 18 eV, respectively, consistent with fragmentation from more highly charged precursors. Similar high-KER channels have been reported for trications, including triply ionized ethylene at 8 × 10$^{14}$ W/cm$^2$, which produces a peak near 11 eV,\cite{xie2015duration} and triply ionized acetylene, for which the total KER is 15 eV.\cite{ibrahim2014tabletop} 

Using the Coulomb repulsion expression inverted to estimate the charge-separation distance at explosion, we extract mean fragment separations of 2.5, 2.1, and 2.24 Å for the doubly ionized channels producing \ce{H+}, \ce{H2+}, and \ce{H3+}, respectively, and 1.7 and 2.0 Å for the higher-energy channels assigned to trication precursors producing \ce{H} and \ce{H2}, respectively. Because the pulses used here are relatively long, bond stretching can occur during the ionization sequence, which favors enhanced ionization at extended bond lengths.\cite{ivanov1996explosive} The observation of \ce{H+} kinetic energy release extending beyond 40 eV is consistent with fragmentation from a highly charged precursor accessed through sequential ionization via enhanced ionization, potentially promoted by proton migration and the associated transient charge localization. If enhanced ionization in this system is promoted by bond stretching, then employing sub-10 fs pulses that are too short for appreciable stretching should curtail the enhanced-ionization pathway and decrease the yield of very high charge states. We plan to test this expectation directly using our recently commissioned 5 fs laser pulses.

Finally, only \ce{H2+} exhibits a near-zero-KER peak. Although a zero-KER component could suggest formation from the monocation, the laser-intensity dependence in Figure 2 indicates that \ce{H2+} appears predominantly at higher intensities and therefore correlates with dication formation. We therefore attribute the zero-KER \ce{H2+} signal to ionization of neutral \ce{H2} (ionization energy 15.4 eV) after it dissociates from the monocation within the same 65 fs laser pulse, before the departing fragments acquire significant relative kinetic energy.
\par
\par
Previous work has shown that the production of \ce{H3+} from methyl pseudohalogens and halogens (\ce{CH3X}) via the \ce{H2} roaming mechanism\cite{ekanayake2017mechanisms,ekanayake2018h2} is governed by three key factors.\cite{stamm2025factors} These are summarized below:

\begin{enumerate}
\item Double ionization leads to a significant elongation of two C–H bonds compared to the neutral CH$_3$X molecule.
\item In the lowest singlet state of CH$_3$X$^{2+}$, the distance between the two hydrogen atoms with the elongated bonds should approximate the ground state \ce{H2} bond length.
\item The adiabatic relaxation energy, defined as the energy difference between the Franck–Condon point (vertical ionization) and the lowest singlet state of the dication, must exceed the dissociation energy of neutral \ce{H2} for its release, a key step in the roaming pathway, but it cannot be so large that other reactive pathways or decomposition mechanisms compete significantly.
\end{enumerate}

\begin{figure}[H]
 \centering
 \includegraphics[width=0.4\textwidth]{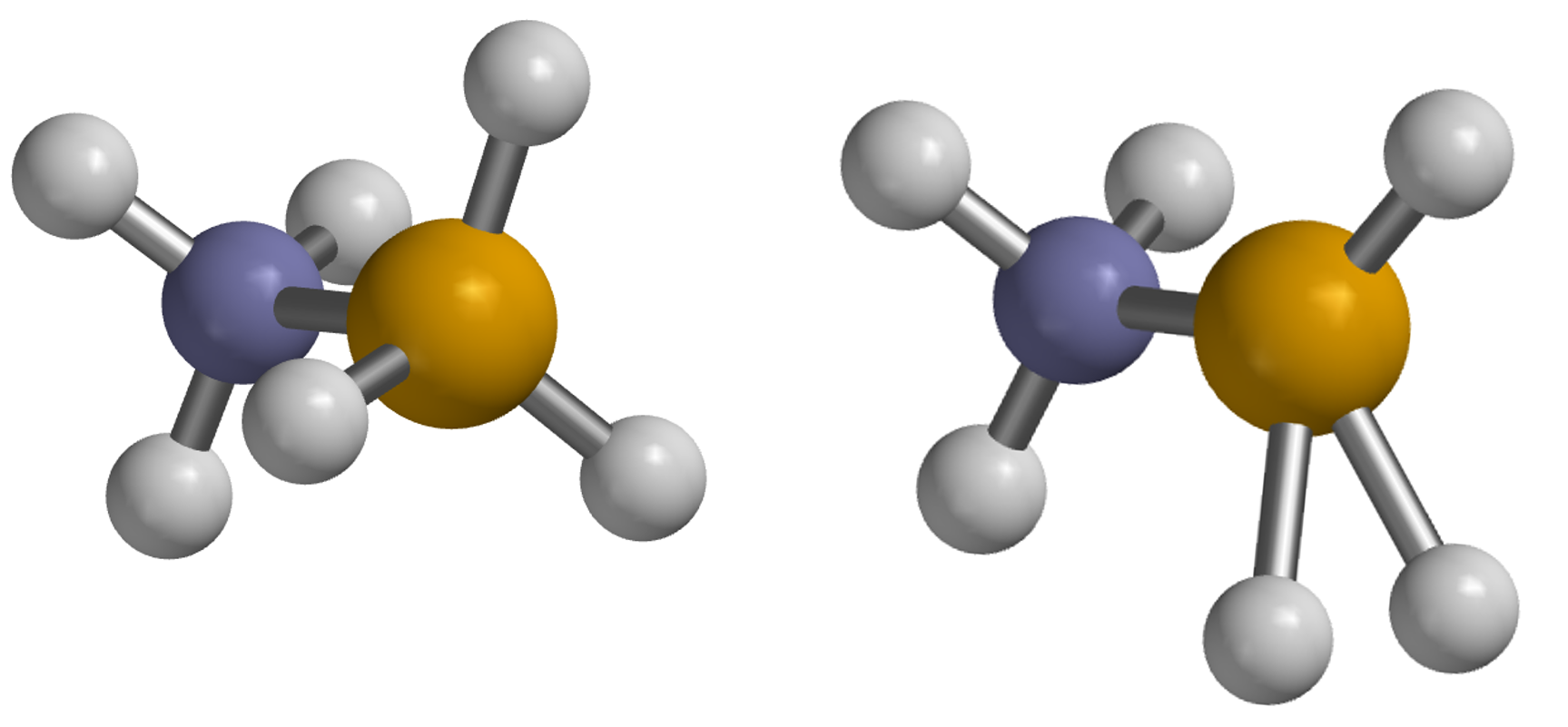}
 \caption{The ground state geometry of the AB neutral (left) and dication (right), visualized using Spartan '24.\cite{spartan2024} The nitrogen atom is blue and the boron atom is tan.}
 \label{Fig6}
\end{figure}

Although AB is not a \ce{CH3X} compound, these governing principles may extend to other molecules, such as ethane,\cite{stamm2025factors} which upon double ionization undergoes H-atom migration and the \ce{CH2CH4^2+} species releases \ce{H2} which roams and abstracts a proton to form \ce{H3+}.\cite{li2020control} To determine if AB, which is isoelectronic with ethane, fulfills these requirements, electronic structure calculations were carried out using GAMESS 2019.R1\cite{GAMESS} at the CCSD(T)\cite{piecuch2002efficient}/aug-cc-pVDZ\cite{dunning1989gaussian,kendall1992electron} level of theory. Comparing the ground state and dication geometries of AB, shown in Figure \ref{Fig6}, one notices the significantly elongated B–H bonds relative to neutral AB, thus satisfying the first criterion.

Additionally, the hydrogens involved in the elongated bonds of the lowest singlet state of AB$^{2+}$ have an internuclear distance of 0.80 \text{\AA} comparable to the 0.74 \text{\AA} bond length of molecular \ce{H2}, thus fulfilling the second condition. The adiabatic relaxation energy of AB$^{2+}$ (7.17 eV) is much higher than the \ce{H2} dissociation energy (1.23 eV), similar to \ce{CH3F^{2+}} (4.5 and 2.2 eV, respectively).\cite{stamm2025factors} The larger energy difference in AB$^{2+}$ favors neutral \ce{H2} release and results in a low \ce{H3+} yield, similar to \ce{CH3F^{2+}}.\cite{newton1970occurrence} Nevertheless, the observation of \ce{H3+} from AB indicates that the same governing factors extend beyond methyl pseudohalogens and halogens, enabling a framework for predicting when \ce{H2} release will lead to appreciable \ce{H3+} formation and providing additional insight into the mechanisms of \ce{H2} release and subsequent \ce{H3+} production in AB.
\par
We performed ab initio molecular dynamics (MD) simulations for both the singly and doubly charged cations of \ce{BH_3NH_3}, using 300 independent trajectories conserving the number of particles, volume, and energy. In all cases, the initial atomic positions and velocities were sampled from a 300 K MD simulation of the neutral parent molecule, so the maximum energy available for fragmentation corresponds to the adiabatic relaxation energy following ionization.

\begin{figure}[h!]

 \includegraphics[width=\columnwidth]{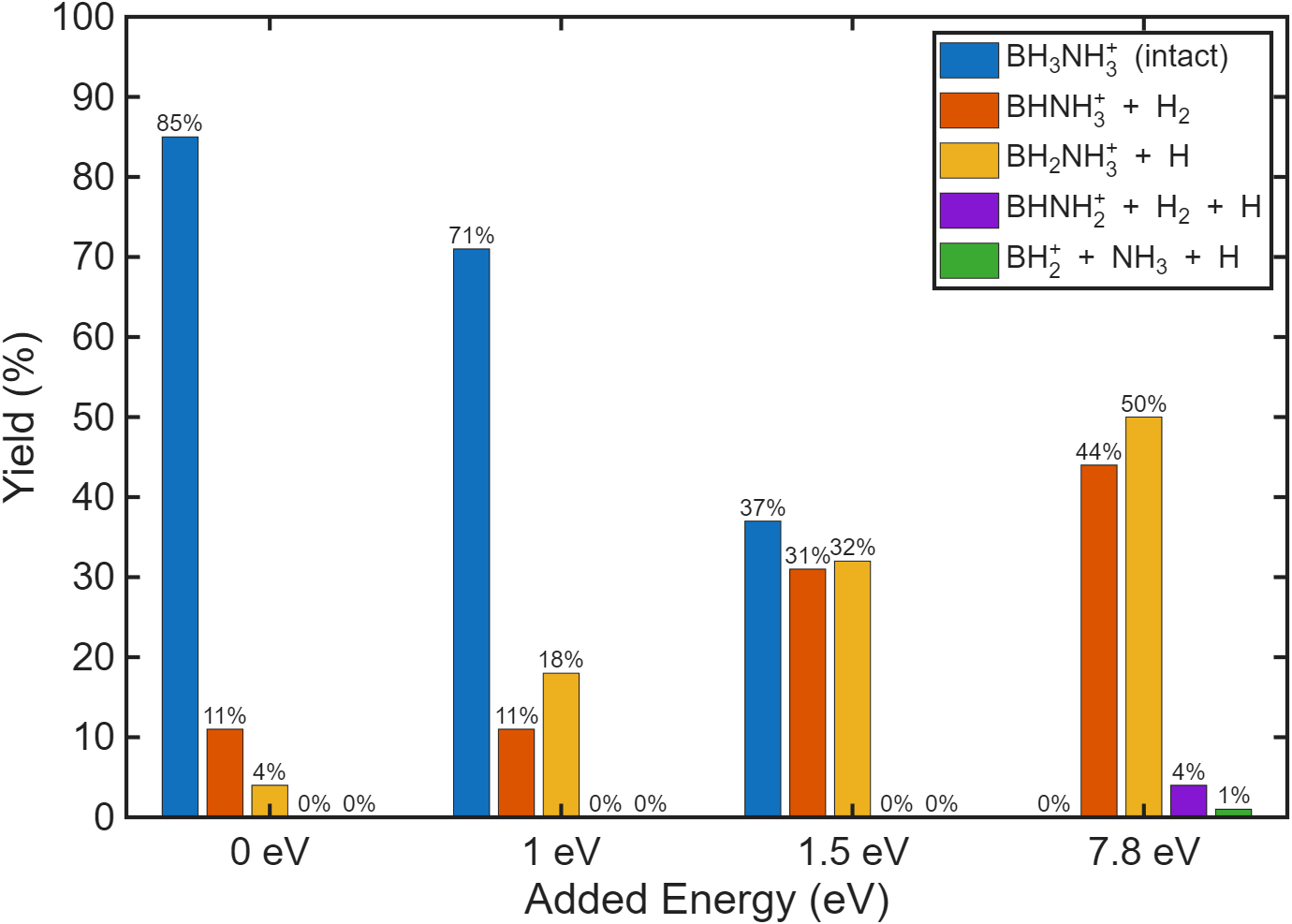}
 \caption{The AIMD trajectory results for the monocation for 0, 1, 1.5, and 7.8 eV kinetic energy.}
 \label{Fig7}
\end{figure}

Under these conditions, the singly charged cation rarely undergoes hydrogen loss, likely due to the 0.74 eV H-bond dissociation energy.\cite{schleier2022ammonia} AIMD trajectories of the monocation were performed with total nuclear kinetic energies of 0, 1, 1.5, and 7.8 eV. 7.8 eV was chosen because the energy difference between the $1e_1$ state and the ground state of the AB cation is 7.76 eV. The 1 and 1.5 eV trajectories were extended to 2 ps in order to capture fragmentation pathways occurring on longer timescales. The results of monocation AIMD trajectories are summarized in Figure \ref{Fig7}. As the total kinetic energy increases, the yield of intact AB decreases while H and/or \ce{H_2} loss becomes more prominent. This dissociation originates exclusively from the boron end of the molecule up to 7.8 eV of deposited energy, consistent with the lower bond dissociation energy of B–H relative to N–H bonds. The increased yield of m/q 30, corresponding to the loss of one H, with increasing internal energy is consistent with the high experimental yield observed for m/q 30. 

Furthermore, we performed AIMD trajectories for the dication (see Movie 1 for a \ce{H3+} formation trajectory,  Movie 2 for a B-N bond cleavage trajectory, Movie 3 for a \ce{H3+} formation trajectory following H scrambling, and Figure S5 for snapshots of these trajectories). After 1~ps, 95\% of the \ce{BH_3NH_3^{2+}} trajectories exhibit hydrogen loss, with product yields following the trend:\[ \ce{H + H^+} > \ce{H_2^+} > \ce{H_2} > \ce{H^+} > \ce{H_3^+} \]
(see Table~\ref{tab:table4}). Following hydrogen loss, \ce{BHNH_3^{2+}} is the most relevant dication for subsequent fragmentation.

The average kinetic energy of nascent \ce{BHNH_3^{2+}} is 1.7~eV, which is insufficient to cleave the B-N bond (dissociation energy $\sim$3.7~eV, see Figure S6). However, this barrier can be overcome experimentally through laser-induced electron recollision. At the laser intensities used here, the recollision energy can exceed 50 eV, which rapidly redistributes as electronic and vibrational excitation. To simulate such conditions and explore B-N bond dissociation, we performed additional MD trajectories for vibrationally excited dications. In these simulations, vibrational energy was injected by either: (i) distributing 5 or 7~eV of kinetic energy among all atoms according to a Boltzmann distribution, or (ii) selectively allocating 4~eV into the normal mode with the largest B-N stretch character (using quasi-classical sampling\cite{hase1977}). 
\begin{table*}[hb]
  \centering
  \begin{tabular}{|c|c|c|}
  \hline
     \textbf{Channel} & \textbf{Yield} & \textbf{Time [fs]}\\
     \hline
         \ce{BH_3NH_3^{2+} \rightarrow \text{intact}} & \(5\%\) & - \\
     \ce{BH_3NH_3^{2+} -> BHNH_3^+ + H + H^+} & \(33\%\) & 73 $\pm$ 55 \\
     \ce{BH_3NH_3^{2+} -> BHNH_3^+ + H_2^+}  & \(23\%\) &134 $\pm$ 74 \\
     \ce{BH_3NH_3^{2+} -> BHNH_3^{2+} + H_2} & \(19\%\) &194 $\pm$ 120 \\
     \ce{BH_3NH_3^{2+} -> BH_2NH_3^+ + H^+}  & \(15\%\) & 47 $\pm$ 23\\
     \ce{BH_3NH_3^{2+} -> BNH_3^+ + H_3^+}  & \(0.5\%\) & - \\
     \ce{BH_3NH_3^{2+} -> BHNH_2^+ + H_3^+}  & \(3.5\%\) &299 $\pm$ 127 \\
  \hline
  \end{tabular}
  \caption{Primary hydrogen-release channels from \ce{BH_3NH_3^{2+}} observed from ab initio molecular dynamics (MD) simulations. Initial atomic positions and velocities were sampled from a 300 K MD trajectory of neutral \ce{BH_3NH_3}. After ionization, trajectories of the dication were propagated in the NVE ensemble.}
  \label{tab:table4}
\end{table*}

\begin{figure*}[ht]
 \centering
 \includegraphics[width=0.7\textwidth]{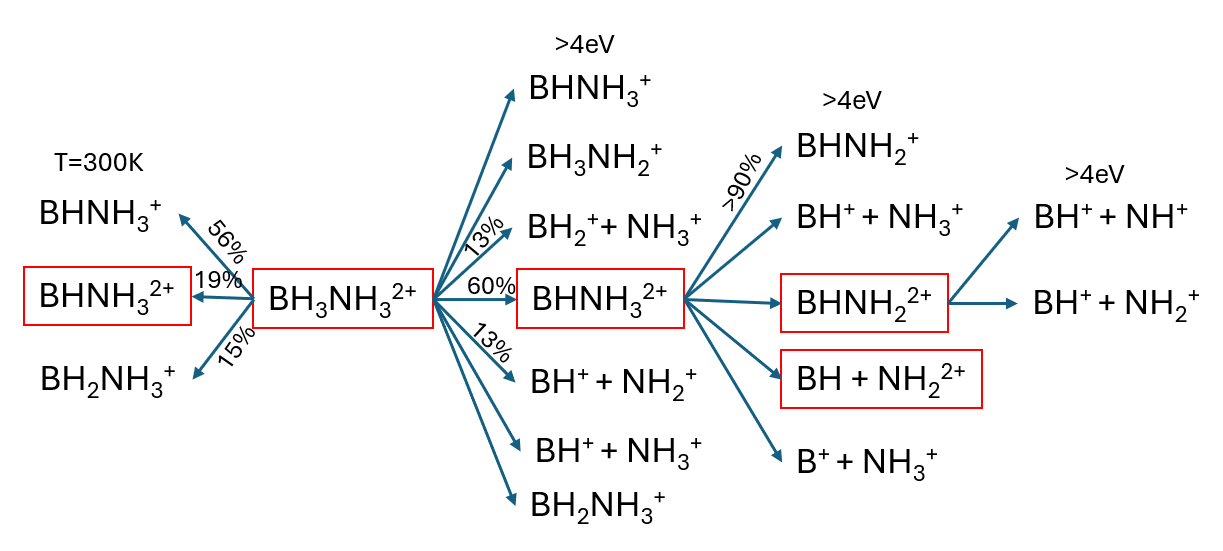}
 \caption{Fragmentation pathways of the doubly charged cation \ce{BH3NH3^{2+}} (red rectangles) following sequential dissociation events. Values accompanying arrows indicate the fraction of trajectories resulting in each specific fragment or set of fragments. Released H's are not indicated for clarity. Numbers on top of each column denote the minimum vibrational energy (in eV) injected into the parent cation during the initial momentum assignment. The leftmost column corresponds to a purely thermal sampling at 300 K, with no additional vibrational energy. Only the most abundant products and primary pathways are shown.}
 \label{pathways}
 
\end{figure*}
The resulting fragmentation pathways are summarized in Figure \ref{pathways}, which shows that H loss in the form of \ce{H+}, \ce{H}, \ce{H2+}, and \ce{H2} are predominantly released from the boron atom. This is in agreement with the positive correlation signal observed between \ce{H+}, \ce{H2+}, and \ce{H3+} and \ce{NH_n^+} (n = 1–3). Even when 4 eV are injected into the vibrational mode that stretches the B–N bond, \ce{H2} is formed approximately 60$\%$ of the time. 
The molecular dynamics simulations support the following assignments for the strong-field ionization mass spectrum: \textit{i)} most abundant dication, $(M-2H)^{2+}$, corresponds to \ce{BHNH3^{2+}}.\textit{ii)} Most abundant cation, $(M-3H)^{+}$, corresponds to \ce{BHNH2^{+}}. \textit{iii)} \ce{NH3+} and \ce{NH2+} can be formed from Coulomb explosion through two channels, while  \ce{NH+} is only significantly observed from fragmentation of \ce{BNH2^{2+}}. This explains why \ce{NH3+} is as abundant as \ce{NH2+} and more than \ce{NH+}.

In general, we never observe a trajectory resulting in \ce{BH3+}, which suggests that the peak in the mass spectrum at $m/q=14$ primarily corresponds to \ce{N+}. Works on the catalytic dehydrogenation of neutral AB show that it is easier to abstract H from B than N,\cite{echeverri2018} consistent with our findings. Summarizing, MD suggests that upon double ionization, \ce{BH3NH3^{2+}} undergoes H-atom/\ce{H2}-molecule elimination prior to Coulomb explosion yielding \ce{BH+}$>$\ce{BH3+}, and \ce{NH3+}$\sim$\ce{NH2+}$>$\ce{NH+} fragments. Also, under single ionization, atomic and molecular hydrogen are most likely released as neutral species.

\par
\section{Conclusions}
In conclusion, we performed time-resolved strong-field ionization of ammonia borane and analyzed its ultrafast fragmentation dynamics. Hydrogen release was observed from both monocationic and dicationic species and occurs on sub-picosecond timescales. Through disruptive probing, we identified that the loss of two and four hydrogen atoms, corresponding to m/q 29 and 27, originates from a common fragmentation pathway and proceeds sequentially. The released hydrogen species predominantly originate from the boron site. Notably, the monocation produces only neutral \ce{H} and \ce{H2}. Fragment correlation analysis identifies the boron center as the primary source of \ce{H+}, \ce{H2+}, and \ce{H3+}. The data obtained also enables the determination of their respective kinetic energy release distributions.

High-level electronic structure calculations were used to determine the first and second ionization potentials of AB, as well as the adiabatic relaxation energy and geometry of the lowest singlet state of the dication. These confirm the AB dication satisfies the structural and energetic criteria for \ce{H2} release, a prerequisite for forming \ce{H3+}. However, the high adiabatic energy causes most \ce{H2} molecules to leave instead of roaming and abstracting a proton. These results extend principles established for the formation of \ce{H3+} via the \ce{H2} roaming mechanism for methyl halogens and pseudohalogens to AB, improving our understanding of far-from-equilibrium chemical processes under strong-field conditions.

AIMD trajectories of the dication reveal extensive \ce{H} scrambling and \ce{H3+} formation via a neutral roaming \ce{H2} intermediate, while also showing that B–N bond cleavage requires substantially higher energy; consistent with experiment, as the earliest \ce{H} loss is initiated from the borane end of the molecule. Given the high hydrogen content of AB and its importance as a promising chemical hydrogen storage material, understanding its gas-phase ionization and fragmentation dynamics provides valuable insight into its dissociative behavior and potential applications.

\section{Supporting Information}
The Supporting Information is available free of charge at: [DOI] It contains: (1) Details about how the threshold laser intensity proportional to the appearance energies were estimated (S-1); (2) Comparison of high and low Intensity dynamics for m/q 29 (S-3); (3) Average of 39,660 mass spectra of \ce{AB} used for calculating the correlation matrix (S-4); (4) Resulting correlation matrix analysis without shot-to-shot corrections (S-5); (5) AIMD Movie snapshots for three trajectories (S-7); (6) Cuts along the potential energy surface as a function of B–N bond distance (S-8); and (7) Table with primary hydrogen-release channels from \ce{AB} (S-9).

\section{Acknowledgements}
This material is based on work supported by the Air Force Office of Scientific Research under award number FA9550-21-1-0428 for the study of ultrafast dynamics induced by secondary electrons. S.K. acknowledges partial funding from the U.S. Department of Energy, Office of Science, Office of Basic Energy Sciences, Atomic, Molecular, and Optical Sciences Program, under SISGER DE-SC0002325. C.C. acknowledges ANID for the grants FONDECYT 1220369 and CEDENNA CIA250002. S.K. acknowledges S. Priyadarsini for her help in setting up the GAMESS calculations. Powered@NLHPC: This research was partially supported by the supercomputing infrastructure of the NLHPC (CCSS210001).

\bibliography{references}

@article{xie2015duration,
  title={Duration of an intense laser pulse can determine the breakage of multiple chemical bonds},
  author={Xie, Xinhua and L{\"o}tstedt, Erik and Roither, Stefan and Sch{\"o}ffler, Markus and Kartashov, Daniil and Midorikawa, Katsumi and Baltu{\v{s}}ka, Andrius and Yamanouchi, Kaoru and Kitzler, Markus},
  journal={Scientific reports},
  volume={5},
  number={1},
  pages={12877},
  year={2015},
  publisher={Nature Publishing Group UK London}
}

@article{ivanov1996explosive,
  title={Explosive ionization of molecules in intense laser fields},
  author={Ivanov, M and Seideman, T and Corkum, P and Ilkov, F and Dietrich, P},
  journal={Physical Review A},
  volume={54},
  number={2},
  pages={1541},
  year={1996},
  publisher={APS}
}

@article{ibrahim2014tabletop,
  title={Tabletop imaging of structural evolutions in chemical reactions demonstrated for the acetylene cation},
  author={Ibrahim, Heide and Wales, Benji and Beaulieu, Samuel and Schmidt, Bruno E and Thir{\'e}, Nicolas and Fowe, Emmanuel P and Bisson, {\'E}ric and Hebeisen, Christoph T and Wanie, Vincent and Gigu{\'e}re, Mathieu and others},
  journal={Nature communications},
  volume={5},
  number={1},
  pages={4422},
  year={2014},
  publisher={Nature Publishing Group UK London}
}

@article{strasser2023single,
  title={Single-photon double-ionisation coulomb explosion in organic molecules},
  author={Strasser, Daniel and Livshits, Ester and Baer, Roi},
  journal={International Reviews in Physical Chemistry},
  volume={42},
  number={1-4},
  pages={29--51},
  year={2023},
  publisher={Taylor \& Francis}
}

@article{jochim2022ultrafast,
  title={Ultrafast disruptive probing: Simultaneously keeping track of tens of reaction pathways},
  author={Jochim, Bethany and DeJesus, Lindsey and Dantus, Marcos},
  journal={Review of Scientific Instruments},
  volume={93},
  number={3}, 
pages ={033003},
  year={2022},
  publisher={AIP Publishing}
}

@article{guo1998single,
  title={Single and double ionization of diatomic molecules in strong laser fields},
  author={Guo, Chunlei and Li, Ming and Nibarger, John P and Gibson, George N},
  journal={Physical Review A},
  volume={58},
  number={6},
  pages={R4271},
  year={1998},
  publisher={APS}
}

@article{dantus2024ultrafast,
  title={Ultrafast studies of elusive chemical reactions in the gas phase},
  author={Dantus, Marcos},
  journal={Science},
  volume={385},
  number={6709},
  pages={eadk1833},
  year={2024},
  publisher={American Association for the Advancement of Science}
}

@article{dantus2024tracking,
  title={Tracking Molecular Fragmentation in Electron--Ionization Mass Spectrometry with Ultrafast Time Resolution},
  author={Dantus, Marcos},
  journal={Accounts of Chemical Research},
  volume={57},
  number={6},
  pages={845--854},
  year={2024},
  publisher={ACS Publications}
}

@article{sa2023photoemission,
  title={Photoemission and X-ray Absorption Investigation of Ammonia-Borane in the Gas Phase},
  author={Sa’adeh, Hanan and Mohamed, Awad EA and Richter, Robert and Coreno, Marcello and Wang, Feng and Prince, Kevin C},
  journal={ACS omega},
  volume={8},
  number={48},
  pages={45970--45975},
  year={2023},
  publisher={ACS Publications}
}

@article{miranda2007ab,
  title={Ab initio investigation of ammonia-borane complexes for hydrogen storage},
  author={Miranda, Caetano R and Ceder, Gerbrand},
  journal={The Journal of chemical physics},
  volume={126},
  number={18},
  year={2007},
  publisher={AIP Publishing}
}

@article{rizzi2019onset,
  title={The onset of dehydrogenation in solid ammonia borane: An ab initio metadynamics study},
  author={Rizzi, Valerio and Polino, Daniela and Sicilia, Emilia and Russo, Nino and Parrinello, Michele},
  journal={Angewandte Chemie International Edition},
  volume={58},
  number={12},
  pages={3976--3980},
  year={2019},
  publisher={Wiley Online Library}
}

@article{castilla2022ammonia,
  title={Ammonia borane-based reactive mixture for trapping and converting carbon dioxide},
  author={Castilla-Martinez, Carlos A and Co{\c{s}}kuner F{\i}l{\i}z, Bilge and Petit, Eddy and Kant{\"u}rk F{\i}gen, Aysel and Demirci, Umit B},
  journal={Frontiers of Materials Science},
  volume={16},
  number={2},
  pages={220610},
  year={2022},
  publisher={Springer}
}

@misc{DOE_H2Storage_ExecSummaries,
  title        = {Executive Summaries for the Hydrogen Storage Materials Centers of Excellence},
  year         = {2012},
  url          = {http://www1.eere.energy.gov/hydrogenandfuelcells/pdfs/executive_summaries_h2_storage_coes.pdf},
  note         = {Accessed: 2025-03-26},
  author       = {{U.S. Department of Energy}}
}

@article{oka2013interstellar,
  title={Interstellar H3+},
  author={Oka, Takeshi},
  journal={Chemical Reviews},
  volume={113},
  number={12},
  pages={8738--8761},
  year={2013},
  publisher={ACS Publications}
}

@article{stamm2025factors,
  title={Factors governing H 3+ formation from methyl halogens and pseudohalogens},
  author={Stamm, Jacob and Priyadarsini, Swati S and Sandhu, Shawn and Chakraborty, Arnab and Shen, Jun and Kwon, Sung and Sandhu, Jesse and Wicka, Clayton and Mehmood, Arshad and Levine, Benjamin G and others},
  journal={Nature Communications},
  volume={16},
  number={1},
  pages={410},
  year={2025},
  publisher={Nature Publishing Group UK London}
}

@article{ekanayake2017mechanisms,
  title={Mechanisms and time-resolved dynamics for trihydrogen cation (H3+) formation from organic molecules in strong laser fields},
  author={Ekanayake, Nagitha and Nairat, Muath and Kaderiya, Balram and Feizollah, Peyman and Jochim, Bethany and Severt, Travis and Berry, Ben and Pandiri, Kanaka Raju and Carnes, Kevin D and Pathak, Shashank and others},
  journal={Scientific reports},
  volume={7},
  number={1},
  pages={4703},
  year={2017},
  publisher={Nature Publishing Group UK London}
}

@article{ekanayake2018h2,
  title={H2 roaming chemistry and the formation of H3+ from organic molecules in strong laser fields},
  author={Ekanayake, Nagitha and Severt, Travis and Nairat, Muath and Weingartz, Nicholas P and Farris, Benjamin M and Kaderiya, Balram and Feizollah, Peyman and Jochim, Bethany and Ziaee, Farzaneh and Borne, Kurtis and others},
  journal={Nature communications},
  volume={9},
  number={1},
  pages={5186},
  year={2018},
  publisher={Nature Publishing Group UK London}
}

@article{ekanayake2018substituent,
  title={Substituent effects on H3+ formation via H2 roaming mechanisms from organic molecules under strong-field photodissociation},
  author={Ekanayake, Nagitha and Nairat, Muath and Weingartz, Nicholas P and Michie, Matthew J and Levine, Benjamin G and Dantus, Marcos},
  journal={The Journal of Chemical Physics},
  volume={149},
  number={24},
  year={2018},
  publisher={AIP Publishing}
}

@article{michie2019quantum,
  title={Quantum coherent control of H3+ formation in strong fields},
  author={Michie, Matthew J and Ekanayake, Nagitha and Weingartz, Nicholas P and Stamm, Jacob and Dantus, Marcos},
  journal={The Journal of Chemical Physics},
  volume={150},
  number={4},
  year={2019},
  publisher={AIP Publishing}
}

@article{kwon2023mechanism,
  title={What is the Mechanism of H3+ Formation from Cyclopropane?},
  author={Kwon, Sung and Sandhu, Shawn and Shaik, Moaid and Stamm, Jacob and Sandhu, Jesse and Das, Rituparna and Hetherington, Caitlin V and Levine, Benjamin G and Dantus, Marcos},
  journal={The Journal of Physical Chemistry A},
  volume={127},
  number={41},
  pages={8633--8638},
  year={2023},
  publisher={ACS Publications}
}

@article{burrows1979studies,
  title={Studies of H+, H+ 2, and H+ 3 dissociative ionization fragments from methane, ethane, methanol, ethanol, and some deuterated methanols using electron-impact excitation and a time-of-flight method incorporating mass analysis},
  author={Burrows, MD and Ryan, SR and Lamb Jr, WE and McIntyre Jr, LC},
  journal={The Journal of Chemical Physics},
  volume={71},
  number={12},
  pages={4931--4940},
  year={1979},
  publisher={American Institute of Physics}
}

@article{hoshina2011metastable,
  title={Metastable decomposition and hydrogen migration of ethane dication produced in an intense femtosecond near-infrared laser field},
  author={Hoshina, Kennosuke and Kawamura, Haruna and Tsuge, Masashi and Tamiya, Minoru and Ishiguro, Masaji},
  journal={The Journal of chemical physics},
  volume={134},
  number={6},
  year={2011},
  publisher={AIP Publishing}
}

@article{kraus2011unusual,
  title={Unusual mechanism for H3+ formation from ethane as obtained by femtosecond laser pulse ionization and quantum chemical calculations},
  author={Kraus, Peter M and Schwarzer, Martin C and Schirmel, Nora and Urbasch, Gunter and Frenking, Gernot and Weitzel, Karl-Michael},
  journal={The Journal of chemical physics},
  volume={134},
  number={11},
  year={2011},
  publisher={AIP Publishing}
}

@article{schirmel2013formation,
  title={Formation of fragment ions (H+, H 3+, CH 3+) from ethane in intense femtosecond laser fields--from understanding to control},
  author={Schirmel, Nora and Reusch, Nicola and Horsch, Philipp and Weitzel, Karl-Michael},
  journal={Faraday Discussions},
  volume={163},
  pages={461--474},
  year={2013},
  publisher={Royal Society of Chemistry}
}

@article{li2020control,
  title={Control of electron recollision and molecular nonsequential double ionization},
  author={Li, Shuai and Sierra-Costa, Diego and Michie, Matthew J and Ben-Itzhak, Itzik and Dantus, Marcos},
  journal={Communications Physics},
  volume={3},
  number={1},
  pages={35},
  year={2020},
  publisher={Nature Publishing Group UK London}
}

@article{piecuch2002efficient,
  title={Efficient computer implementation of the renormalized coupled-cluster methods: the r-ccsd [t], r-ccsd (t), cr-ccsd [t], and cr-ccsd (t) approaches},
  author={Piecuch, Piotr and Kucharski, Stanis{\l}aw A and Kowalski, Karol and Musia{\l}, Monika},
  journal={Computer Physics Communications},
  volume={149},
  number={2},
  pages={71--96},
  year={2002},
  publisher={Elsevier}
}

@article{GAMESS,
title = {Recent developments in the general atomic and molecular electronic structure system},
volume = {152},
issn = {0021-9606, 1089-7690},
url = {http://aip.scitation.org/doi/10.1063/5.0005188},
doi = {10.1063/5.0005188},
language = {en},
number = {15},
urldate = {2020-06-18},
journal = {The Journal of Chemical Physics},
author = {Barca, Giuseppe M. J. and Bertoni, Colleen and Carrington, Laura and Datta, Dipayan and De Silva, Nuwan and Deustua, J. Emiliano and Fedorov, Dmitri G. and Gour, Jeffrey R. and Gunina, Anastasia O. and Guidez, Emilie and Harville, Taylor and Irle, Stephan and Ivanic, Joe and Kowalski, Karol and Leang, Sarom S. and Li, Hui and Li, Wei and Lutz, Jesse J. and Magoulas, Ilias and Mato, Joani and Mironov, Vladimir and Nakata, Hiroya and Pham, Buu Q. and Piecuch, Piotr and Poole, David and Pruitt, Spencer R. and Rendell, Alistair P. and Roskop, Luke B. and Ruedenberg, Klaus and Sattasathuchana, Tosaporn and Schmidt, Michael W. and Shen, Jun and Slipchenko, Lyudmila and Sosonkina, Masha and Sundriyal, Vaibhav and Tiwari, Ananta and Galvez Vallejo, Jorge L. and Westheimer, Bryce and Wloch, Marta and Xu, Peng and Zahariev, Federico and Gordon, Mark S.},
month = apr,
year = {2020},
pages = {154102}
}

@article{luque2025anomalous,
  title={Anomalous Ionization in the Central Molecular Zone by Sub-GeV Dark Matter},
  author={Luque, Pedro De la Torre and Balaji, Shyam and Silk, Joseph},
  journal={Physical Review Letters},
  volume={134},
  number={10},
  pages={101001},
  year={2025},
  publisher={APS}
}

@article{herbst1973formation,
  title={The formation and depletion of molecules in dense interstellar clouds},
  author={Herbst, Eric and Klemperer, William},
  journal={The Astrophysical Journal},
  volume={185},
  pages={505--534},
  year={1973}
}

@article{graetz2009new,
  title={New approaches to hydrogen storage},
  author={Graetz, Jason},
  journal={Chemical Society Reviews},
  volume={38},
  number={1},
  pages={73--82},
  year={2009},
  publisher={Royal Society of Chemistry}
}

@article{chen2008recent,
  title={Recent progress in hydrogen storage},
  author={Chen, Ping and Zhu, Min},
  journal={Materials today},
  volume={11},
  number={12},
  pages={36--43},
  year={2008},
  publisher={Elsevier}
}

@misc{spartan2024,
  title        = {Spartan'24, Version 1.2.0},
  author       = {{Wavefunction Inc.}},
  year         = {2024},
  howpublished = {\url{https://www.wavefun.com}},
  note         = {Irvine, CA, USA}
}

@article{corkum1993plasma,
  title={Plasma perspective on strong field multiphoton ionization},
  author={Corkum, Paul B},
  journal={Physical review letters},
  volume={71},
  number={13},
  pages={1994},
  year={1993},
  publisher={APS}
}

@article{posthumus2004dynamics,
  title={The dynamics of small molecules in intense laser fields},
  author={Posthumus, JH},
  journal={Reports on Progress in Physics},
  volume={67},
  number={5},
  pages={623},
  year={2004},
  publisher={IOP Publishing}
}

@article{wang2005disentangling,
  title={Disentangling the volume effect through intensity-difference spectra: Application to laser-induced dissociation of H 2+},
  author={Wang, Pengqian and Sayler, A Max and Carnes, Kevin D and Esry, Brett D and Ben-Itzhak, Itzik},
  journal={Optics letters},
  volume={30},
  number={6},
  pages={664--666},
  year={2005},
  publisher={Optical Society of America}
}

@article{wiese2019strong,
  title={Strong-field photoelectron momentum imaging of OCS at finely resolved incident intensities},
  author={Wiese, Joss and Olivieri, Jean-Fran{\c{c}}ois and Trabattoni, Andrea and Trippel, Sebastian and K{\"u}pper, Jochen},
  journal={New Journal of Physics},
  volume={21},
  number={8},
  pages={083011},
  year={2019},
  publisher={IOP Publishing}
}

@article{schleier2022ammonia,
  title={Ammonia Borane, NH3BH3: A Threshold Photoelectron--Photoion Coincidence Study of a Potential Hydrogen-Storage Material},
  author={Schleier, Domenik and Gerlach, Marius and Pratim Mukhopadhyay, Deb and Karaev, Emil and Schaffner, Dorothee and Hemberger, Patrick and Fischer, Ingo},
  journal={Chemistry--A European Journal},
  volume={28},
  number={42},
  pages={e202201378},
  year={2022},
  publisher={Wiley Online Library}
}

@article{lozovoy2008control,
  title={Control of molecular fragmentation using shaped femtosecond pulses},
  author={Lozovoy, Vadim V and Zhu, Xin and Gunaratne, Tissa C and Harris, D Ahmasi and Shane, Janelle C and Dantus, Marcos},
  journal={The Journal of Physical Chemistry A},
  volume={112},
  number={17},
  pages={3789--3812},
  year={2008},
  publisher={ACS Publications}
}

@article{newton1970occurrence,
  title={The occurrence of the H3+ ion in the mass spectra of organic compounds},
  author={Newton, Amos S and Sciamanna, AF and Thomas, GE},
  journal={International Journal of Mass Spectrometry and Ion Physics},
  volume={5},
  number={5-6},
  pages={465--482},
  year={1970},
  publisher={Elsevier}
}

@article{echeverri2018,
   author = {Echeverri, Andrea and Cárdenas, Carlos and Calatayud, Monica and Hadad, Cacier Zilahy and Gomez, Tatiana},
   title = {Theoretical analysis of the adsorption of Ammonia-Borane and their dehydrogenation products on the (001) surface of TiC and ZrC},
   journal = {Surface Science},
   ISSN = {0039-6028},
   DOI = {doi.org/10.1016/j.susc.2018.10.016},
   url = {https://doi.org/10.1016/j.susc.2018.10.016},
   year = {2018},
   type = {Journal Article}
}

@article{hase1977,
   author = {Sloane, Christine S. and Hase, William L.},
   title = {On the dynamics of state selected unimolecular reactions: Chloroacetylene dissociation and predissociation},
   journal = {The Journal of Chemical Physics},
   volume = {66},
   number = {4},
   pages = {1523-1533},
   ISSN = {0021-9606
1089-7690},
   DOI = {10.1063/1.434116},
   year = {1977},
   type = {Journal Article}
}

@misc{gaussian09,
   author = {Frisch, M. J. and Trucks, G. W. and Schlegel, H. B. and Scuseria, G. E. and Robb, M. A. and Cheeseman, J. R. and Scalmani, G. and Barone, V. and Mennucci, B. and Petersson, G. A. and Nakatsuji, H. and Caricato, M. and Li, X. and Hratchian, H. P. and Izmaylov, A. F. and Bloino, J. and Zheng, G. and Sonnenberg, J. L. and Hada, M. and Ehara, M. and Toyota, K. and Fukuda, R. and Hasegawa, J. and Ishida, M. and Nakajima, T. and Honda, Y. and Kitao, O. and Nakai, H. and Vreven, T. and Montgomery, J. A. and Peralta, J. E. and Ogliaro, F. and Bearpark, M. and Heyd, J. J. and Brothers, E. and Kudin, K. N. and Staroverov, V. N. and Kobayashi, R. and Normand, J. and Raghavachari, K. and Rendell, A. and Burant, J. C. and Iyengar, S. S. and Tomasi, J. and Cossi, M. and Rega, N. and Millam, J. M. and Klene, M. and Knox, J. E. and Cross, J. B. and Bakken, V. and Adamo, C. and Jaramillo, J. and Gomperts, R. and Stratmann, R. E. and Yazyev, O. and Austin, A. J. and Cammi, R. and Pomelli, C. and Ochterski, J. W. and Martin, R. L. and Morokuma, K. and Zakrzewski, V. G. and Voth, G. A. and Salvador, P. and Dannenberg, J. J. and Dapprich, S. and Daniels, A. D. and Farkas and Foresman, J. B. and Ortiz, J. V. and Cioslowski, J. and Fox, D. J.},
   title = {Gaussian 09, Revision B.01},
   url = {},
   year = {2009},
   type = {Generic}
}

@article{hase1999,
   author = {Millam, John M. and Bakken, Vebjo/rn and Chen, Wei and Hase, William L. and Schlegel, H. Bernhard},
   title = {Ab initio classical trajectories on the Born–Oppenheimer surface: Hessian-based integrators using fifth-order polynomial and rational function fits},
   journal = {The Journal of Chemical Physics},
   volume = {111},
   number = {9},
   pages = {3800-3805},
   ISSN = {0021-9606},
   DOI = {10.1063/1.480037},
   url = {https://doi.org/10.1063/1.480037},
   year = {1999},
   type = {Journal Article}
}

@article{w97xd,
   author = {Chai, Jeng-Da and Head-Gordon, Martin},
   title = {Long-range corrected hybrid density functionals with damped atom–atom dispersion corrections},
   journal = {Phys. Chem. Chem. Phys.},
   volume = {10},
   number = {44},
   pages = {6615-6},
   DOI = {10.1039/b810189b},
   url = {https://doi.org/10.1039/b810189b},
   year = {2008},
   type = {Journal Article}
}

@article{piecuch2005renormalized,
  title={Renormalized coupled-cluster methods exploiting left eigenstates of the similarity-transformed Hamiltonian},
  author={Piecuch, Piotr and W{\l}och, Marta},
  journal={The Journal of chemical physics},
  volume={123},
  number={22},
  year={2005},
  publisher={AIP Publishing}
}

@article{shen2013doubly,
  title={Doubly electron-attached and doubly ionized equation-of-motion coupled-cluster methods with 4-particle--2-hole and 4-hole--2-particle excitations and their active-space extensions},
  author={Shen, Jun and Piecuch, Piotr},
  journal={The Journal of chemical physics},
  volume={138},
  number={19},
  year={2013},
  publisher={AIP Publishing}
}

@article{shen2014doubly,
  title={Doubly electron-attached and doubly ionised equation-of-motion coupled-cluster methods with full and active-space treatments of 4-particle--2-hole and 4-hole--2-particle excitations: The role of orbital choices},
  author={Shen, Jun and Piecuch, Piotr},
  journal={Molecular Physics},
  volume={112},
  number={5-6},
  pages={868--885},
  year={2014},
  publisher={Taylor \& Francis}
}

@article{vcivzek1966correlation,
  title={On the correlation problem in atomic and molecular systems. Calculation of wavefunction components in Ursell-type expansion using quantum-field theoretical methods},
  author={{\v{C}}{\'\i}{\v{z}}ek, Ji{\v{r}}{\'\i}},
  journal={The Journal of Chemical Physics},
  volume={45},
  number={11},
  pages={4256--4266},
  year={1966},
  publisher={American Institute of Physics}
}

@article{vcivzek1969use,
  title={On the use of the cluster expansion and the technique of diagrams in calculations of correlation effects in atoms and molecules},
  author={{\v{C}}{\'\i}{\v{z}}ek, Ji{\v{r}}{\'\i}},
  journal={Advances in chemical physics},
  pages={35--89},
  year={1969},
  publisher={Wiley Online Library}
}

@article{paldus1972correlation,
  title={Correlation Problems in Atomic and Molecular Systems. IV. Extended Coupled-Pair Many-Electron Theory and Its Application to the B H 3 Molecule},
  author={Paldus, J and {\v{C}}{\'\i}{\v{z}}ek, J and Shavitt, I},
  journal={Physical Review A},
  volume={5},
  number={1},
  pages={50},
  year={1972},
  publisher={APS}
}

@article{purvis1982full,
  title={A full coupled-cluster singles and doubles model: The inclusion of disconnected triples},
  author={Purvis III, George D and Bartlett, Rodney J},
  journal={The Journal of chemical physics},
  volume={76},
  number={4},
  pages={1910--1918},
  year={1982},
  publisher={American Institute of Physics}
}

@article{cullen1982linked,
  title={The linked singles and doubles model: An approximate theory of electron correlation based on the coupled-cluster ansatz},
  author={Cullen, John M and Zerner, Michael C},
  journal={The Journal of Chemical Physics},
  volume={77},
  pages={4088--4109},
  year={1982},
  publisher={American Institute of Physics}
}

@article{Eland1996TheOrigin,
author = {Eland, J. H. D.},
title = {The Origin of Primary H},
journal = {Rapid Communications in Mass Spectrometry},
volume = {10},
pages = {1560-1562},
year = {1996}
}

@article{furukawa2005ejection,
  title={Ejection of triatomic hydrogen molecular ion from methanol in intense laser fields},
  author={Furukawa, Yusuke and Hoshina, Kennosuke and Yamanouchi, Kaoru and Nakano, Hidetoshi},
  journal={Chemical physics letters},
  volume={414},
  number={1-3},
  pages={117--121},
  year={2005},
  publisher={Elsevier}
}

@article{okino2006ejection,
  title={Ejection dynamics of hydrogen molecular ions from methanol in intense laser fields},
  author={Okino, T and Furukawa, Y and Liu, P and Ichikawa, T and Itakura, R and Hoshina, K and Yamanouchi, K and Nakano, H},
  journal={Journal of Physics B: Atomic, Molecular and Optical Physics},
  volume={39},
  number={13},
  pages={S515},
  year={2006},
  publisher={IOP Publishing}
}

@article{gope2022inverse,
  title={An “inverse” harpoon mechanism},
  author={Gope, Krishnendu and Livshits, Ester and Bittner, Dror M and Baer, Roi and Strasser, Daniel},
  journal={Science Advances},
  volume={8},
  number={39},
  pages={eabq8084},
  year={2022},
  publisher={American Association for the Advancement of Science}
}

@article{gope2023sequential,
  title={Sequential mechanism in H 3+ formation dynamics on the ethanol dication},
  author={Gope, Krishnendu and Bittner, Dror M and Strasser, Daniel},
  journal={Physical Chemistry Chemical Physics},
  volume={25},
  number={9},
  pages={6979--6986},
  year={2023},
  publisher={Royal Society of Chemistry}
}

@article{raghavachari1989fifth,
  title={A fifth-order perturbation comparison of electron correlation theories},
  author={Raghavachari, Krishnan and Trucks, Gary W and Pople, John A and Head-Gordon, Martin},
  journal={Chemical Physics Letters},
  volume={157},
  number={6},
  pages={479--483},
  year={1989},
  publisher={Elsevier}
}

@article{bartlett1990non,
  title={Non-iterative fifth-order triple and quadruple excitation energy corrections in correlated methods},
  author={Bartlett, Rodney J and Watts, JD and Kucharski, SA and Noga, J},
  journal={Chemical physics letters},
  volume={165},
  number={6},
  pages={513--522},
  year={1990},
  publisher={Elsevier}
}

@article{dunning1989gaussian,
  title={Gaussian basis sets for use in correlated molecular calculations. I. The atoms boron through neon and hydrogen},
  author={Dunning Jr, Thom H},
  journal={The Journal of chemical physics},
  volume={90},
  number={2},
  pages={1007--1023},
  year={1989},
  publisher={American Institute of Physics}
}

@article{woon1993gaussian,
  title={Gaussian basis sets for use in correlated molecular calculations. III. The atoms aluminum through argon},
  author={Woon, David E and Dunning Jr, Thom H},
  journal={The Journal of chemical physics},
  volume={98},
  number={2},
  pages={1358--1371},
  year={1993},
  publisher={American Institute of Physics}
}

@article{kendall1992electron,
  title={Electron affinities of the first-row atoms revisited. Systematic basis sets and wave functions},
  author={Kendall, Rick A and Dunning Jr, Thom H and Harrison, Robert J},
  journal={The Journal of chemical physics},
  volume={96},
  number={9},
  pages={6796--6806},
  year={1992},
  publisher={American Institute of Physics}
}

@article{kumar2021reduction,
  title={Reduction of carbon dioxide with ammonia-borane under ambient conditions: maneuvering a catalytic way},
  author={Kumar, Abhishek and Eyyathiyil, Jusaina and Choudhury, Joyanta},
  journal={Inorganic Chemistry},
  volume={60},
  number={15},
  pages={11684--11692},
  year={2021},
  publisher={ACS Publications}
}

@article{frasinski1989covariance,
  title={Covariance mapping: A correlation method applied to multiphoton multiple ionization},
  author={Frasinski, LJ and Codling, K and Hatherly, PA},
  journal={Science},
  volume={246},
  number={4933},
  pages={1029--1031},
  year={1989},
  publisher={American Association for the Advancement of Science}
}

@article{li2024fragment,
  title={Fragment Correlation Mass Spectrometry Enables Direct Characterization of Disulfide Bond Cleavage Pathways of Therapeutic Peptides},
  author={Li, Yangjie and Cavet, Guy and Zare, Richard N and Driver, Taran},
  journal={Analytical Chemistry},
  volume={96},
  number={38},
  pages={15081--15084},
  year={2024},
  publisher={ACS Publications}
}

@article{li2024fragment1,
  title={Fragment correlation mass spectrometry: Determining the structures of biopolymers in a complex mixture without isolating individual components},
  author={Li, Yangjie and Cavet, Guy and Zare, Richard N and Driver, Taran},
  journal={Proceedings of the national academy of sciences},
  volume={121},
  number={32},
  pages={e2409676121},
  year={2024},
  publisher={National Academy of Sciences}
}

@article{de2000femtosecond,
  title={Femtosecond Dynamics of Norrish Type-II Reactions: Nonconcerted Hydrogen-Transfer and Diradical Intermediacy},
  author={De Feyter, Steven and Diau, Eric W-G and Zewail, Ahmed H},
  journal={Angewandte Chemie},
  volume={112},
  number={1},
  pages={266--269},
  year={2000},
  publisher={Wiley Online Library}
}

@article{stamm2023surprising,
  title={The surprising dynamics of the mclafferty rearrangement},
  author={Stamm, Jacob and Kwon, Sung and Sandhu, Shawn and Shaik, Moaid and Das, Rituparna and Sandhu, Jesse and Curenton, Bradley and Wicka, Clayton and Levine, Benjamin G and Sun, Liangliang and others},
  journal={The Journal of Physical Chemistry Letters},
  volume={14},
  number={44},
  pages={10088--10093},
  year={2023},
  publisher={ACS Publications}
}

@article{minvielle2025isomer,
  title={How Isomer and Conformer Structures Impact Dissociation Dynamics of Alkane Radical Cations},
  author={Minvielle, Madison K and Aftel, Mikaela and Hill, Timothy and Lopez Pena, Hugo A and Tibbetts, Katharine Moore},
  journal={The Journal of Physical Chemistry A},
  volume={129},
  number={37},
  pages={8550--8562},
  year={2025},
  publisher={ACS Publications}
}

@article{sutton2011regeneration,
  title={Regeneration of ammonia borane spent fuel by direct reaction with hydrazine and liquid ammonia},
  author={Sutton, Andrew D and Burrell, Anthony K and Dixon, David A and Garner III, Edward B and Gordon, John C and Nakagawa, Tessui and Ott, Kevin C and Robinson, J Pierce and Vasiliu, Monica},
  journal={Science},
  volume={331},
  number={6023},
  pages={1426--1429},
  year={2011},
  publisher={American Association for the Advancement of Science}
}

\end{document}


\maketitle

\renewcommand*\contentsname{Table of Contents}
\renewcommand{\baselinestretch}{1.0}\normalsize 
{\small \tableofcontents} 
\renewcommand{\baselinestretch}{1.0}\normalsize 

\newpage

\section{Appearance Energy Calculation}
Included here is an illustration of the procedure used to extract appearance intensities from the intensity-dependent fragment yields. For each fragment, the integrated yield is converted to a difference trace by subtracting point nn from point n+1n+1, which suppresses focal-volume contributions. The resulting trace is fit with an error function, and the onset intensity is obtained from the fitted curve (arrow shown for m/qm/q 18). This same workflow is applied to the major fragments summarized in the intensity-dependent heat map in Figure 2 of the main text.
\begin{figure*}[h!]
  \centering
  \includegraphics[width=0.5
  \textwidth]{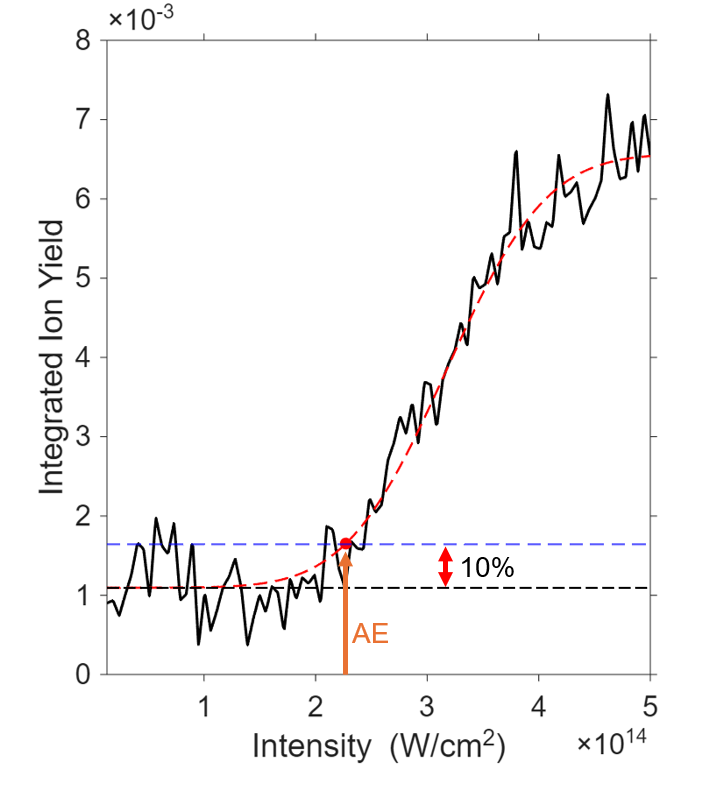}
  \caption{The integrated ion yield of m/q 18 as a function of laser intensity. The black line corresponds to the experimental data while the red dashed line is the error function fit. The orange arrow indicates where the appearance energy of m/q 18 is.}
  \label{FigS1}
\end{figure*}
\newpage

\section{Comparison of High and Low Intensity Dynamics}
The following figure is included to demonstrate that pump–probe transients recorded at nominally high intensity can be dominated by contributions from lower-intensity regions of the interaction volume when no intensity-selection procedure is applied. We test this explicitly using the m/q 29 channel (loss of two H atoms): the delay-dependent ion yield measured at $3.1 \times 10^{14}$ W/cm$^2$ closely matches the transient measured at $1.1 \times 10^{14}$ W/cm$^2$. The near overlap of these traces supports the conclusion that the apparent high-intensity dynamics are strongly weighted by lower-intensity contributions, motivating the high-intensity selection approach used in the main analysis.

\begin{figure*}[h]
  \centering
  \includegraphics[width=0.5
  \textwidth]{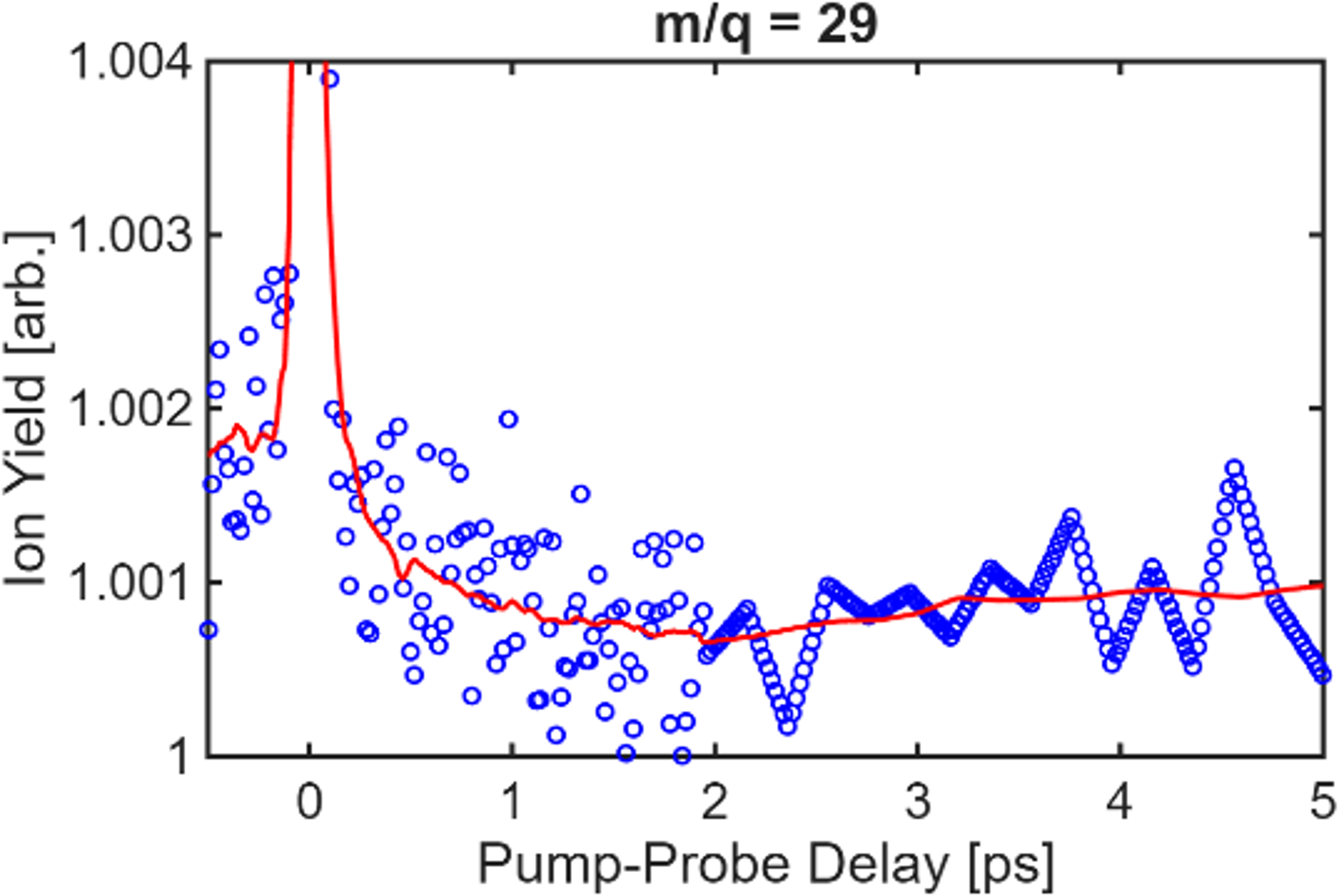}
  \caption{Comparison of ion yields for m/q 29 as a function of pump–probe delay at high intensity (red line, $3.1 \times 10^{14}$ W/cm$^2$) and low intensity (blue circles, $1.1 \times 10^{14}$ W/cm$^2$).}
  \label{FigS2}
\end{figure*}
\newpage

\section{Mass Spectrum of AB for Correlation Coefficient Analysis}
Here we highlight the quality and key features of the mass spectrum used for the fragment correlation and kinetic energy analyses. The figure shows a representative portion of the average of 39,660 single-shot mass spectra acquired at $3.1 \times 10^{14}$ W/cm$^2$, highlighting that many peaks in the m/q 2–17 region exhibit distinct forward and backward components characteristic of Coulomb explosion. These split features provide the basis for assigning charge-state-specific channels and are subsequently used in the fragment correlation analysis (Figure 4) and the KER analysis (Figure 5) in the main text.
\begin{figure*}[h]
  \centering
  \includegraphics[width=0.8
  \textwidth]{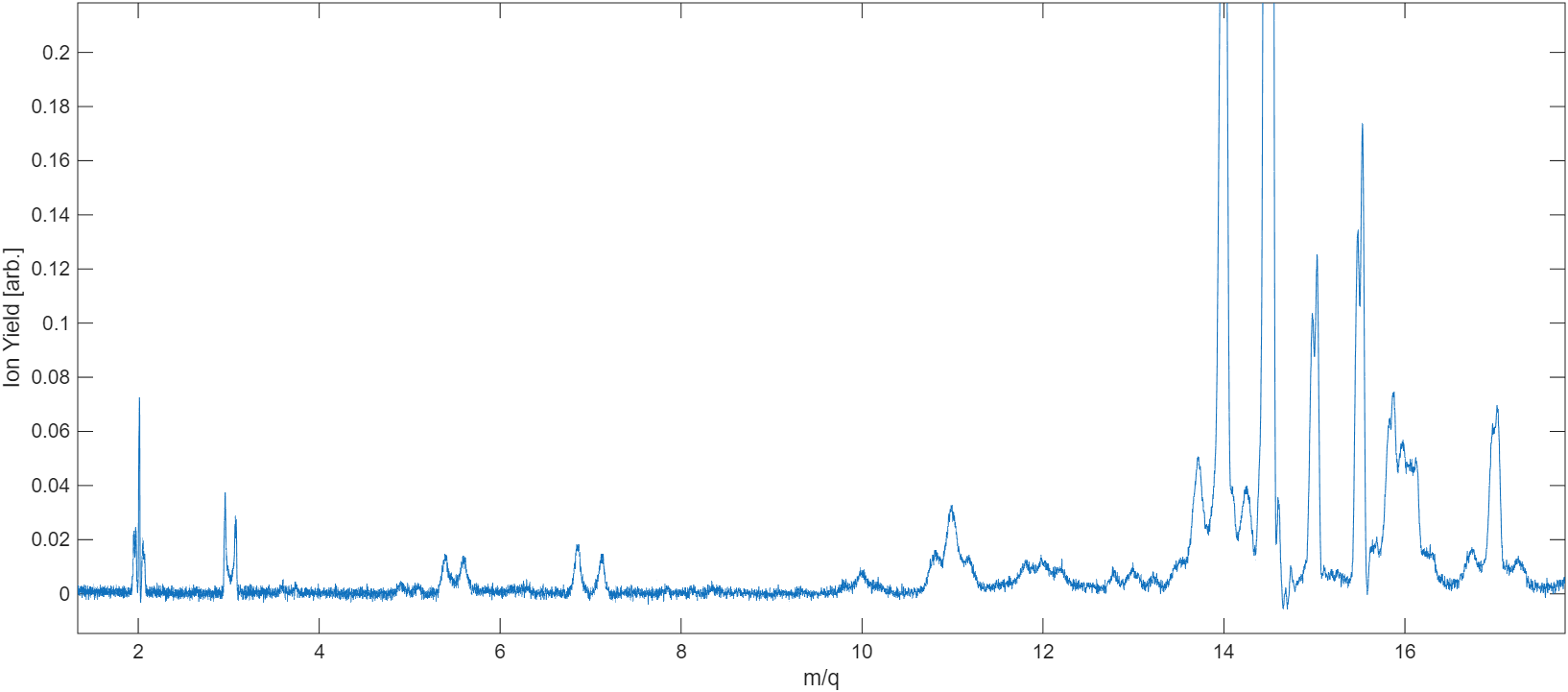}
  \caption{Zoomed in MS of AB from m/q 2 to 17.}
  \label{FigS3}
\end{figure*}
\newpage

\section{Correlation Matrix without Corrections}
Shot-to-shot fluctuations in the total ion signal due to variations in laser intensity introduce a strong common-mode contribution to all fragment yields. Without normalizing for these fluctuations, nearly all m/q channels appear positively correlated, producing an almost uniform correlation background that masks pathway-specific relationships. This matrix, shown in Figure S4,  therefore illustrates why fluctuation correction is essential: otherwise, the correlations are dominated by experimental variability rather than mechanistic links such as common parentage or concerted breakup.
\begin{figure*}[h]
  \centering
  \includegraphics[width=0.6
  \textwidth]{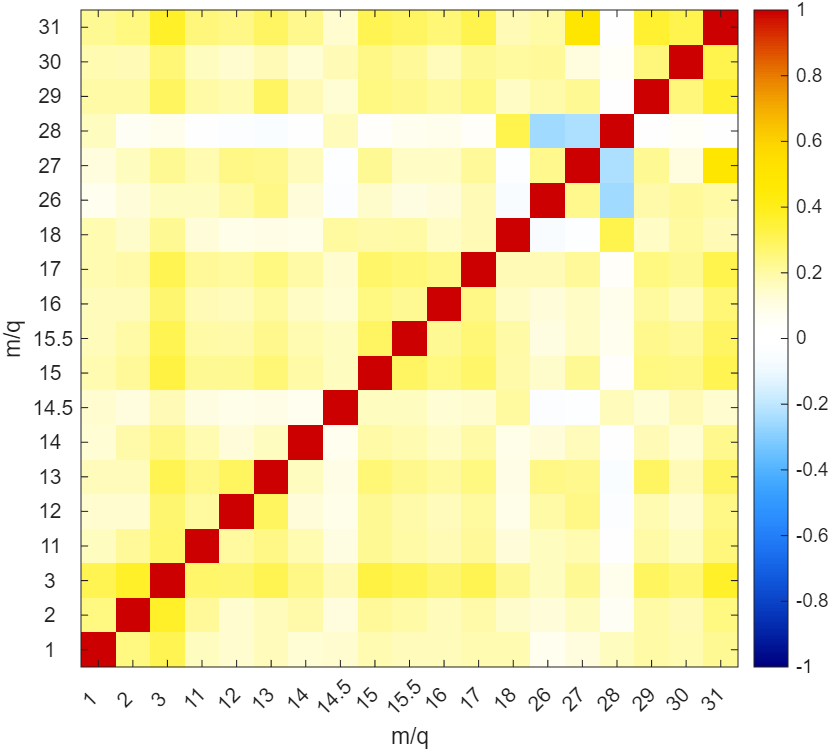}
  \caption{Correlation matrix of AB fragments before correcting for shot-to-shot fluctuations}
  \label{FigS4}
\end{figure*}
\newpage

\section{AIMD Movie Snapshots}
To complement the experimental fragment-correlation analysis, we performed ab initio molecular dynamics (AIMD) simulations on doubly ionized ammonia borane \ce{AB2+}
to directly visualize the ultrafast structural rearrangements and dissociation pathways accessible after double ionization. Representative trajectory snapshots are presented below, highlighting \ce{H3+} formation, \ce{B}–\ce{N} bond cleavage, and \ce{H3+} formation following hydrogen scrambling (see Movies 1–3).
\begin{figure*}[h]
  \centering
  \includegraphics[width=1.1
  \textwidth]{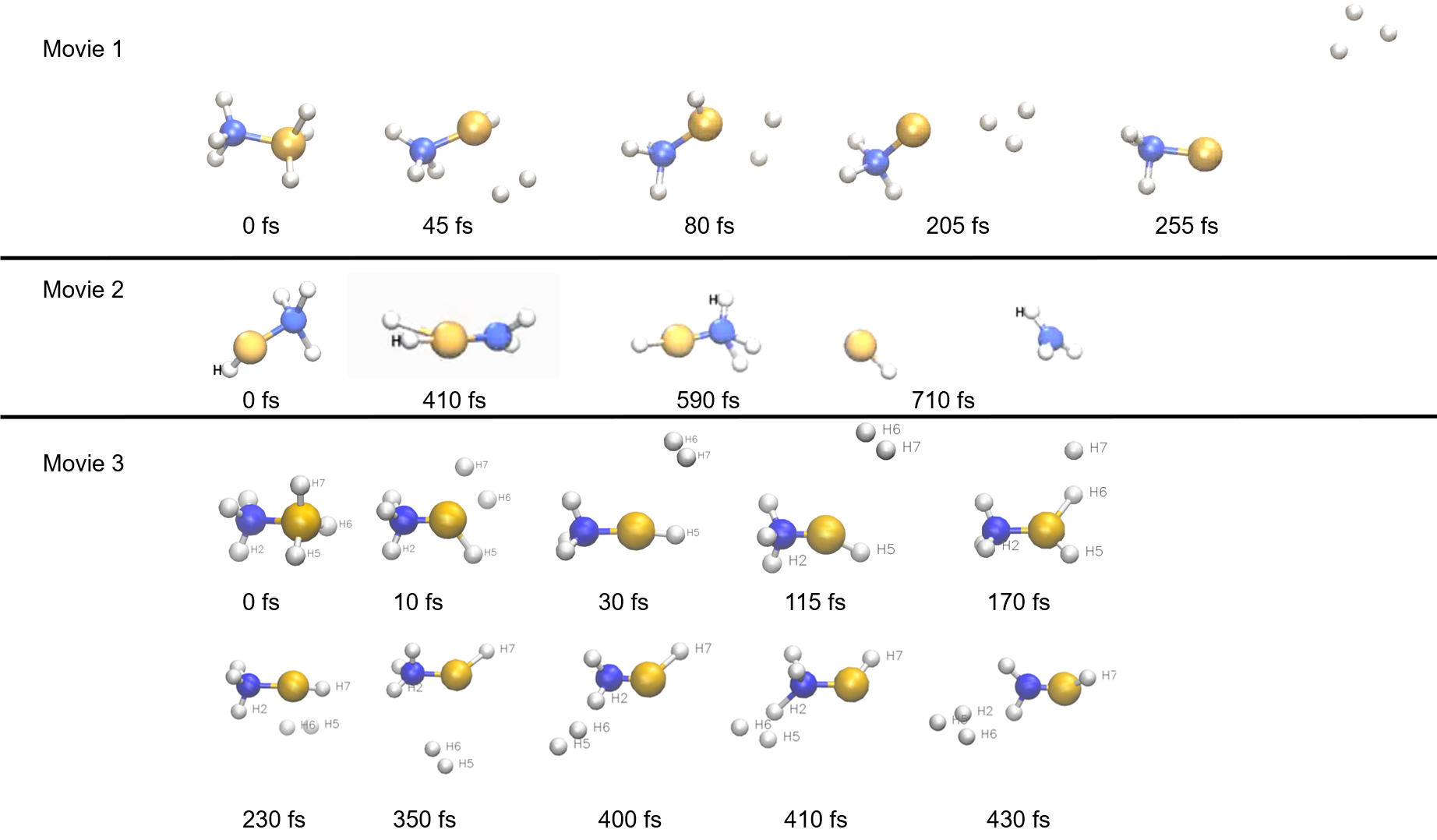}
  \caption{The ab initio molecular dynamics trajectory snapshots show the formation of \ce{H3+} from doubly charged ammonia borane (AB), B–N bond cleavage following double ionization, and \ce{H3+} formation from doubly charged AB after hydrogen scrambling.
}
  \label{FigS5}
\end{figure*}
\newpage

\section{B-N Bond Breaking Energy}
To help interpret the bond-cleavage energetics, we include one-dimensional cuts of the potential energy surface along the \ce{B}–\ce{N} stretch coordinate. The plot reports the electronic energy as a function of the \ce{B}–\ce{N} distance, with each curve referenced to its own ground-state minimum (zero energy). For \ce{BH3NH3+}, the curve exhibits a discontinuity near 2.9 Å that reflects a change in the dominant reaction coordinate: as the \ce{B}–\ce{N} bond is elongated, one hydrogen migrates from boron to nitrogen (not shown), leading to formation of the separated products \ce{BH2} and \ce{NH4+} rather than simple homolytic \ce{B}–\ce{N} bond dissociation.

\begin{figure*}[h!]
  \centering
  \includegraphics[width=0.8\textwidth]{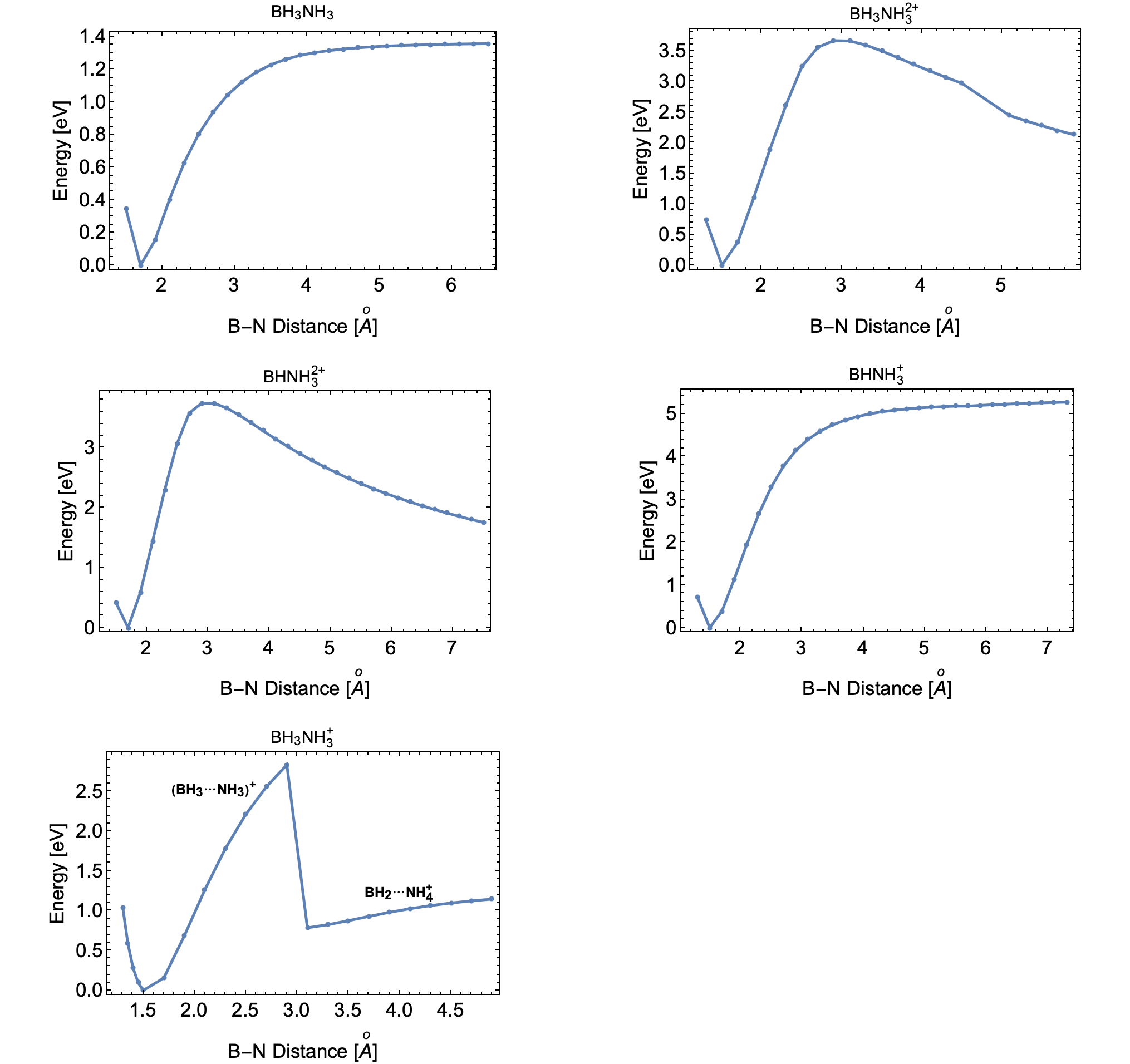}
  \caption{Potential energy as a function of the B-N distance. The zero energy is set to the ground state energy of each compound. In the case of \ce{BH3NH3+}, the discontinuity around 2.9 \AA \space corresponds to the migration of one of the H from B to N (not shown) to form the products \ce{BH2} and \ce{NH4+}.}
  \label{FigS6}
\end{figure*}
\newpage

\section{Primary hydrogen-release channels from \ce{BH_3NH_3^+}}

\begin{table*}[h]
    \centering
    \begin{tabular}{|c|c|}
    \hline
         \textbf{Channels \ce{BH_3NH_3^+}} & \textbf{Yield} \\
         \hline
                  \ce{BH_3NH_3^{+} \rightarrow \text{intact}}  & \(85\%\) \\
         \ce{BH_3NH_3^{+} -> BHNH_3^+ + H }           & \(4\%\) \\
         \ce{BH_3NH_3^{+} -> BHNH_3^{+} + H_2}        & \(11\%\)\\
         \hline
         \textbf{Channels \ce{BH_3NH_3^+} @ 1 eV} & \textbf{Yield} \\
         \hline
                  \ce{BH_3NH_3^{+} \rightarrow \text{intact}}  & \(71\%\) \\
         \ce{BH_3NH_3^{+} -> BHNH_3^+ + H }           & \(18\%\) \\
         \ce{BH_3NH_3^{+} -> BHNH_3^{+} + H_2}        & \(11\%\)\\
         \hline
         \textbf{Channels \ce{BH_3NH_3^+} @ 1.5 eV} & \textbf{Yield} \\
         \hline
                  \ce{BH_3NH_3^{+} \rightarrow \text{intact}}  & \(37\%\) \\
         \ce{BH_3NH_3^{+} -> BHNH_3^+ + H }           & \(32\%\) \\
         \ce{BH_3NH_3^{+} -> BHNH_3^{+} + H_2}        & \(31\%\)\\
         \hline
         
    \hline
    \end{tabular}
     \caption{Primary hydrogen-release channels from \ce{BH_3NH_3^{+}} observed from ab initio molecular dynamics (MD) simulations. Initial atomic positions were sampled from a 300 K MD trajectory of neutral \ce{BH_3NH_3}. After ionization, trajectories of the cation were propagated in the NVE ensemble.  Velocities in the  NVE ensemble  were thermally  sampled (\ce{BH_3NH_3^+}) or adjusted to distribute 1 and 1.5  eV of kinetic energy among all atoms according to a Boltzmann distribution (\ce{BH_3NH_3^+} @ 1 eV or @ 1.5 eV)}
    \label{tab:table4}
\end{table*}